\documentclass[12pt]{JHEP3}
\usepackage[centertags]{amsmath}
\usepackage{amssymb}
\usepackage{epsfig}
\usepackage{amsmath}
\usepackage{graphicx}
\usepackage{afterpage}

\def\beq{\begin{equation}}
\def\eeq{\end{equation}}
\def\bea{\begin{eqnarray}}
\def\eea{\end{eqnarray}}

\def\nl{\nonumber\\}

\def\roughly#1{\mathrel{\raise.3ex\hbox
{$#1$\kern-.75em\lower1ex\hbox{$\sim$}}}}
\def\lsim{\roughly<}
\def\gsim{\roughly>}
\def\lesssim{\mathrel{\hbox{\rlap{\hbox{\lower4pt\hbox{$\sim$}}}\hbox{$<$}}}}
\def\gtrsim{\mathrel{\hbox{\rlap{\hbox{\lower4pt\hbox{$\sim$}}}\hbox{$>$}}}}

\def\sla#1{\raise.15ex\hbox{$/$}\kern-.57em #1}
\def\bra#1{\left\langle #1\right|}
\def\ket#1{\left| #1\right\rangle}

\def\barr{\begin{eqnarray}}
\def\earr{\end{eqnarray}}
\def\beast{\begin{eqnarray*}}
\def\eeast{\end{eqnarray*}}

\def\be{\begin{equation}}
\def\ee{\end{equation}}
\def\bea{\begin{eqnarray}}
\def\eea{\end{eqnarray}}

\def\A0{{\cal{A}}_{0}}
\def\Apar{{\cal{A}}_{\|}}
\def\Aperp{{\cal{A}}_{\perp}}

\def\AtP{{\cal{A}}_{tP}}

\def\ssV{ \sin {2 \theta_{D^*}}}
\def\sVsq{ \sin^2{\theta_{D^*}}}
\def\cV{ \cos { \theta_{D^*}}}
\def\c2V{ \cos { 2 \theta_{D^*}}}
\def\cVsq{ \cos^2{\theta_{D^*}}}
\def\sl{ \sin { \theta_{l}}}
\def\s2l{ \sin {2 \theta_{l}}}
\def\slsq{ \sin^2{ \theta_{l}}}
\def\cl{ \cos { \theta_{l}}}
\def\c2l{ \cos { 2 \theta_{l}}}
\def\clsq{ \cos^2{ \theta_{l}}}

\def\beq{\begin{equation}}
\def\eeq{\end{equation}}
\def\bea{\begin{eqnarray}}
\def\eea{\end{eqnarray}}

\def\nl{\nonumber\\}
\def\roughly#1{\mathrel{\raise.3ex\hbox
{$#1$\kern-.75em\lower1ex\hbox{$\sim$}}}}
\def\lsim{\roughly<}
\def\gsim{\roughly>}
\def\lesssim{\mathrel{\hbox{\rlap
{\hbox{\lower4pt\hbox{$\sim$}}}\hbox{$<$}}}}
\def\gtrsim{\mathrel{\hbox{\rlap
{\hbox{\lower4pt\hbox{$\sim$}}}\hbox{$>$}}}}

\newcommand{\bcln}{b \to c l^- \bar{\nu}_{l}}

\newcommand{\barBstdtn}{\bar{B} \to D^{*} \tau \bar{\nu}_{\tau}}

\def\BDtaunu{\bar{B} \to D^+ \tau^{-} \bar{\nu_\tau}} 
 
\def\BDlnu{\bar{B} \to D^+ \ell^{-} \bar{\nu_\ell}}
\def\BDstartaunu{\bar{B} \to D^{*+} \tau^{-} \bar{\nu_\tau}}
\def\BDstarellnu{\bar{B} \to D^{*+} l^{-} \bar{\nu_\ell}}

\def\BDstarlnu{\bar{B} \to D^{*} \ell^{-} \bar{\nu_\ell}}

\def\bra#1{\left\langle #1\right|}
\def\ket#1{\left| #1\right\rangle}

\def \thl {{\theta_l}}
\def \thD {{\theta_{D^*}}}
\title{\boldmath The Full $B \to D^{*} \tau^{-} \bar{\nu_\tau}$ Angular Distribution
 and CP violating Triple Products}

\author{
Murugeswaran Duraisamy$^a$ and Alakabha Datta$^a$
\\
$^a$ Department of Physics and Astronomy, 108 Lewis Hall,
\\ ~~University of Mississippi, Oxford, MS 38677-1848, USA 
\\
E-mail:
\email{duraism@phy.olemiss.edu}, \email{datta@phy.olemiss.edu}
}

\abstract{ We perform a comprehensive study of the impact of
  new-physics operators with different Lorentz structures on
$\BDstarellnu$ decays, $(\ell = e, \mu, \tau $)
involving the $b \to c l \nu_\ell$ transition.
We present the full three angle and $q^2$ angular distribution
with new physics operators with complex couplings. 
Various observables are constructed from the angular distribution with
 special focus on the CP violating triple product asymmetries which vanish
in the Standard Model without any hadronic complications. 
Two of the three triple products are only sensitive to  vector/axial vector new physics operators. 
Hence, the measurements of non-zero triple-product asymmetries will be a clear sign of new   physics and a strong signal for vector/axial vector new physics operators. Even though we focus on $\tau$ final state, 
one can use the triple-products to search for new physics  with $e$ and $\mu$ in the final state.  
 
}


\preprint{UMISS-HEP-2013-04}

\begin{document}

\section{Introduction}
The Standard Model (SM) of particle physics, even though very successful, is expected to break down at some energy scale and make way for a more complete theory. Exploration of what lies beyond the SM can be carried out at the energy frontier in colliders such as the LHC or at the intensity frontier at high luminosity experiments. In the intensity frontier, the B factories, BaBar and Belle, have produced an enormous  quantity of data in the last decade. There is still a lot of data to be analyzed from both experiments. 
The B factories have firmly established the CKM mechanism as the leading order contributor to CP violating phenomena in the flavor sector involving quarks. New physics (NP) effects can add to the leading order term producing deviations from the SM predictions.  In this 
respect, the second and third generation quarks and leptons are 
quite special because they are comparatively heavier and are 
expected to be relatively more sensitive to NP. 
As an example, in certain versions of the two Higgs doublet models (2HDM), the couplings of the new Higgs bosons are proportional to the masses and so NP effects are more pronounced for the heavier generations.
Moreover, the constraints on NP  involving the third generation leptons and quarks are somewhat weaker allowing for larger NP effects. 

It is interesting that there are certain discrepancies in decays involving
$\tau$ and $\nu_{\tau}$ states, though none of them are significant enough 
to establish clearly the presence of NP.
There is  a seeming violation of universality in the tau lepton coupling to the W suggested by the Lep II data which could indicate NP associated with the third generation lepton \cite{MartinGon}.
Recent measurement of CP violation \cite {BABAR:2011aa}  in $\tau$ decays
find 
$A_{CP}$ in $\tau^- \to \pi^-K_s( \ge 0 \pi^0) \nu_{\tau}$ is $(-0.36 \pm 0.23 \pm 0.11)$ \%  which is different from the SM prediction
$(0.36  \pm 0.01)$ by 2.8 $\sigma$.
The branching ratio of $B \to \tau \nu_{\tau}$ showed some tension with the SM  predictions \cite{belletau} indicating  NP, possibly coming from an extended scalar or gauge sector \cite{Bhattacherjee:2010ju}.
However,  new  Belle \cite{Adachi:2012mm} and BaBar \cite{Lees:2012ju} measurements,  obtained  using the hadronic tagging method, are more consistent with the SM.


If there is NP involving the third generation leptons one can search for it in semileptonic $ b \to c \tau \nu_{\tau}$ decays such as $\BDtaunu$, $\BDstartaunu$ \cite{nierste}.
The semileptonic decays of B meson to the $\tau$ lepton is 
mediated by a $W$ boson in the SM and it is quite well understood 
theoretically. In many models of NP  this decay gets 
contributions from additional states like new vector bosons or 
new scalar particles. The exclusive decays $\BDtaunu$ 
and $\BDstartaunu$ are important places to look for NP 
because, being three-body decays, they offer a host of observables 
in the angular distributions of the final state particles. The 
theoretical uncertainties of the SM predictions have gone down 
significantly in recent years because of the developments in 
heavy-quark effective theory (HQET). The experimental situation 
has also improved a lot since the first observation of the decay 
$\BDstartaunu$ in 2007 by the Belle Collaboration 
\cite{Matyja:2007kt}. After 2007 many improved measurements have 
been reported by both the BaBar and Belle collaborations and the 
evidence for the decay  $\BDtaunu$ has also been found 
\cite{Aubert:2007dsa,Adachi:2009qg,Bozek:2010xy}. 
Recently, the BaBar collaboration with their full data sample 
of an integrated luminosity of 426 fb$^{-1}$ has reported the measurements 
of the quantities \cite{:2012xj}
\begin{eqnarray}
\label{babarnew}
R(D) &=& \frac{BR(\BDtaunu)}
{BR(\BDlnu)}=0.440 \pm 0.058 \pm 0.042\, ,
\nonumber \\
R(D^*) &=& \frac{BR(\BDstartaunu)}
{BR(\BDstarlnu)}=0.332 \pm 0.024 \pm 0.018 \, .
\end{eqnarray}
The SM predictions for $R(D)$ and $R(D^*)$ are 
\cite{:2012xj,Fajfer:2012vx,Sakaki:2012ft}
\begin{eqnarray}
R(D) &=& 0.297 \pm 0.017 \, ,
\nonumber \\
R(D^*) &=& 0.252 \pm 0.003 \,,
\end{eqnarray}
which deviate from the BaBar measurements by 2$\sigma$ and 2.7$\sigma$ 
respectively. The BaBar collaboration  reported a 3.4$\sigma$ 
deviation from SM when the two measurements of Eq.~(\ref{babarnew}) are taken 
together. 

These deviations could be sign of NP and already certain models of NP have been considered to explain the data \cite{Fajfer:2012vx,Fajfer:2012jt, Crivellin:2012ye,Datta:2012qk,Becirevic:2012jf, Deshpande:2012rr,  Celis:2012dk,Choudhury:2012hn,Tanaka:2012nw,Ko:2012sv,Fan:2013qz,Biancofiore:2013ki,Celis:2013jha}.
In Ref.~\cite{Datta:2012qk}, we calculated various observables in $\BDtaunu$ and $\BDstartaunu$ decays with NP  using an effective Lagrangian approach. 
The Lagrangian contains two quarks and two leptons with scalar, pseudoscalar, vector, axial vector and tensor operators. Considering  the NP operators one at a time, the  coefficient of these operators can be fixed from the BaBar measurements and then one can  study the effect of these operators on the various observables.  In this work, we extend the work of Ref.~\cite{Datta:2012qk}
by providing the full angular distribution for $\BDstartaunu$ with NP. 
The full angular distribution, in the SM, has already been used in experiments for final states with
muon and the electron. 
In this work we also consider CP violating observables which are  the triple product (TP) asymmetries  \cite{TP}.
In the SM, these TPs  rigorously vanish and so any non-zero measurements of these terms are clear signs of NP without any hadronic uncertainties.  In the presence of NP
with complex couplings the TP's are non-zero and  depend on the form factors.
Moreover, as we will see most of the TPs depend on the vector/axial vector couplings
and not on the pseudoscalar couplings. Hence these TPs provide useful clues to the nature of NP. 
As in the previous work, we will neglect the tensor term in the effective Lagrangian.

The paper is organized in the following manner. In Sec. 2   we set up the formalism where we introduce the effective Lagrangian for NP, define the various helicity amplitudes  and consider the constraints on the NP couplings. In Sec. 3 we present the angular distribution  
and define the various observables in  $\BDstartaunu$ decays. We  present the SM  predictions for these observables 
as well as predictions for the  observables with NP.  Finally, in Sec. 4 we summarize the results of our analysis.
\section{Formalism}
In the presence of NP, the effective Hamiltonian for the quark-level transition $\bcln$  can be written in the form \cite{ccLag} 
\bea
{\cal{H}}_{eff} &=& \frac{4 G_F V_{cb}}{\sqrt{2}} \Big[ (1 + V_L)\,[\bar{c} \gamma_\mu P_L b] ~ [\bar{l} \gamma^\mu P_L \nu_l] \, +  V_R \, [\bar{c} \gamma^\mu P_R b] ~ [\bar{l} \gamma_\mu P_L \nu_l] \nl && \, + S_L \, [\bar{c} P_L b] \,[\bar{l}  P_L \nu_l] \, +  S_R \, ~[\bar{c} P_R b] \,~ [\bar{l}  P_L \nu_l]  \,  + T_L \, [\bar{c} \sigma^{\mu \nu} P_L b] \,~[\bar{l} \sigma_{\mu \nu} P_L \nu_l]\Big]\,,
\eea 
where  $G_F = 1.1663787(6) \times 10^{-5} GeV^{-2}$ is the Fermi coupling constant, $V_{cb}$ is the Cabibbo-Koboyashi-Maskawa (CKM) matrix element, $P_{L,R} = ( 1 \mp \gamma_5)/2$  are the projectors of negative/positive chiralities. We use $\sigma_{\mu \nu} = i[\gamma_\mu, \gamma_\nu]/2$ and
assume the neutrino to be always left chiral.
Further, we do not assume any relation between $b \to u l^- \nu_l$ and $\bcln$ transitions and hence  do not include constraints from
$B \to \tau \nu_{\tau}$. 
The SM  effective Hamiltonian corresponds to $V_L = V_R = S_L = S_R =  T_L = T_R = 0$. In this paper we will ignore the  tensor interactions. With this simplification we write the effective Lagrangian as 
\bea
\label{eq1:Lag}
{\cal{H}}_{eff} &=&  \frac{G_F V_{cb}}{\sqrt{2}}\Big\{
\Big[\bar{c} \gamma_\mu (1-\gamma_5) b  + g_V \bar{c} \gamma_\mu  b + g_A \bar{c} \gamma_\mu \gamma_5 b\Big] \bar{l} \gamma^\mu(1-\gamma_5) \nu_l \nl && +  \Big[g_S\bar{c}  b   + g_P \bar{c} \gamma_5 b\Big] \bar{l} (1-\gamma_5)\nu_l + h.c \Big\}, \
\eea
where $g_{V,A} =  V_R \pm V_L$ and $g_{S,P} =  S_R \pm S_L$.\vspace*{1mm}
The values of the couplings that can explain the data in Eq.~(\ref{babarnew}) satisfy the constraints
$|g_{V,A}| \lsim 2 $ and $|g_{P}| \lsim 4 $. One can consider if the size of these couplings can arise in typical extensions of the SM. Let us start with the vector/axial vector couplings and assume that the new physics is due to the exchange of a new particle with mass $M_X$ with coupling $g_{new}$ to the quarks which has the same size as the weak coupling, $g$, of the quarks to the $W$. One can then write
\bea
\frac{g_{new}^2}{8 M_X^2} & \approx & \frac{g^2}{8M_W^2} V_{cb} g_{V,A}. \
\eea
With $g_{new} \approx g $ one obtains,
\bea
g_{V,A} & \approx & \frac{M_W^2}{M_X^2 V_{cb}}.\
\eea
Hence $M_X \approx $ 300 GeV can lead to $g_{V,A} \approx 2$. Note that such a particle
which couples dominantly to the third family is still allowed by experimental searches.
Coming to the pseudoscalar coupling, one notes that the hadronic matrix elements
are somewhat suppressed so larger values of $g_{P}$, satisfying $|g_{P}| \lsim 4 $, are needed to explain the data. In this case $M_X \approx $ 200 GeV can lead to $g_{P} \approx 4$ and such a particle
which couples dominantly to the third family is still allowed by experimental searches.

The effects of NP can be seen in the helicity amplitudes that describe the decays.
The expressions for the  hadronic helicity amplitudes for the $\barBstdtn$ decays are
\bea
\label{tran_amp}
{\cal{A}}_0  &=&\frac{1}{2 m_{D^*} \sqrt{q^2}} \Big[(m_B^2 - m_{D^*}^2 - q^2) (m_B + m_{D^*} ) A_1(q^2) -\frac{4 m_B^2 |p_{D^*}|^2}{m_B +m_{D^*}}  A_2(q^2) \Big](1 - g_A) \,,\nl
{\cal{A}}_{\|}  &=& \sqrt{2}(m_B + m_{D^*})A_1(q^2) (1 - g_A)  \,,\nl
{\cal{A}}_{\perp}  &=&   -\sqrt{2} \frac{2 m_B V(q^2)}{(m_B + m_{D^*})} |p_{D^*}|(1 + g_V) \,,\nl
{\cal{A}}_{t}  &=& \frac{2 m_B |p_{D^*}| A_0(q^2) }{\sqrt{q^2}} (1 - g_A) \,,\nl
{\cal{A}}_{P}  &=& -\frac{2 m_B |p_{D^*}| A_0(q^2) }{ (m_b(\mu) + m_c(\mu))} g_P \,,
\eea
where the $t$ and the $P$ amplitudes arise in the combination
\bea
\label{tp_comb}
{\cal{A}}_{tP} &=& \Big({\cal{A}}_t + \frac{\sqrt{q^2}}{m_\tau} {\cal{A}}_P \Big)\,.
\eea
The form factors $A_{1,2,0}(q^2)$ and $V(q^2)$ are defined in the appendix. As is clear from the above equation, the various helicity amplitudes are sensitive to different NP operators. These helicity amplitudes can be probed in various differential distributions providing useful information about NP.

The transversity amplitudes $\Apar$ and $\Aperp $ are related to the helicity amplitudes $\cal{A}_{\pm}$  as
\bea
\label{tran_basis}
\cal{A}_{\perp} &= &\frac{1}{\sqrt{2}}  \left( \cal{A}_+ - \cal{A}_{-} \right), \nonumber\\
\cal{A}_{\|} &= &\frac{1}{\sqrt{2}}  \left( \cal{A}_+ + \cal{A}_ {-}\right). \
\eea
All the amplitudes are complex if the NP couplings are complex.
The phases in the couplings are weak phases and change sign when we go from particle to anti-particle decays. Though strong phases in the current can arise from higher-order
loops these will be tiny and we will ignore them. Hence the only CP violating signals
will be of the triple-product type and all direct CP violating effects will vanish.
Moreover we see that ${\cal{A}}_{0,\|,t} $ have the same weak phases and any interference between these amplitudes will not lead to any CP violating signals.
The only CP violating signals will come from the interference of ${\cal{A}}_{\perp} $ with the other vector/axial vector and pseudoscalar current amplitudes ${\cal{A}}_{0,\|,tP} $.

We will now consider the two cases:
\begin{itemize}
\item { Case a :  In this case, we will set $S_L, S_R=0$ and assume that the NP affects leptons of only the third generation. This scenario could arise from the exchange of a new charged
$W^{\prime}$ boson \cite{dattaneutrino}. We point out that this is just a simplifying assumption and in fact the general angular distribution presented in the paper is also applicable to  $e$ and $\mu$ in the final state.
}

\item{ Case b :  
In this case, we will set $V_L, V_R=0$ and assume that the NP only affects  leptons of the third generation. This scenario could arise in models with extended scalar sectors \cite{dattaBnp}.
}
\end{itemize}
Finally, we discuss the possibility of long distance resonant contribution to this decay
as one observes in $B \to K^{(*)} \ell^+ \ell^-$ decays. Note that
the decay $\BDstartaunu$ is a tree level decay unlike $B \to K^{(*)} \ell^+ \ell^-$. The decay can
get long distance contributions from $B \to D^* X$ with the subsequent decay $X \to \tau \nu_{\tau}$ which is an annihilation process and is suppressed. 
The state $X$, given the energy required to produce the $\tau$, can be $D_s^{(*)}, D^{(*)}$ e.t.c.
The branching ratio 
for $B \to D^* X$ is smaller than $\BDstartaunu$ and  combined with the suppressed
rate for $X \to \tau \nu_{\tau}$  the resonant long distance contribution in this case is much smaller than the leading tree level $W$ exchange contribution and can be neglected.

\subsection{Constraints on the NP couplings}
\label{constraints}

For the numerical calculation, we use the $B \to D$ and $B \to D^*$ form factors in the heavy quark effective theory framework \cite{Neubert:1993mb,Caprini:1997mu}. $B \to D^*$ form factors are summarized in the appendix. The constraints on the complex NP couplings in  the  $\bcln$ effective Hamiltonian come  from 
the measured $R(D)$ and $R(D^*)$ in Eq.~(\ref{babarnew}) at 95\% C.L.  We also vary the free parameters in the form factors discussed in the appendix  within their error bars. All the other numerical values are taken from \cite{Nakamura:2010zzi} and \cite{Asner:2010qj}.  A detailed analysis of $R(D)$ and NP in the decay $\BDlnu$ can be found in \cite{Datta:2012qk}. The allowed  ranges for the  NP couplings are then used for predicting the allowed ranges  for the observables in the the angular distribution discussed in the next section. The experimental results show a correlation between $R(D)$ and $R(D^*)$. Many NP models would affect both $R(D)$ and $R(D^*)$ and produce a correlation between them, while other NP models would affect only one of the ratios. We believe the measured deviations from the SM for both $R(D)$ and $R(D^*)$ are not significant enough to rule out the SM or NP models that affect only one of the ratios. Hence, in our determination of the allowed ranges of the NP couplings the correlations between $R(D)$ and $R(D^*)$ are not taken into account. In the future, if experiments find  more significant deviations from the SM predictions for the two ratios, or other clear signals for NP in these decays, then the effect of the correlation will have to be taken into account to find the nature of the NP. The goal of the paper is to point out how different observables in these decays can to be used to find NP and the nature of the NP.

\begin{figure}[h!]
\centering
\includegraphics[width=6.5cm,height=5.5cm]{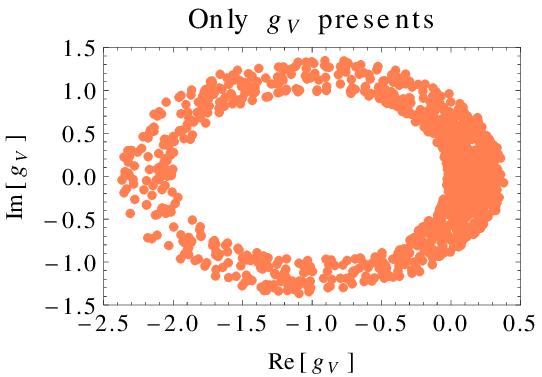}\,
\includegraphics[width=6.5cm,height=5.5cm]{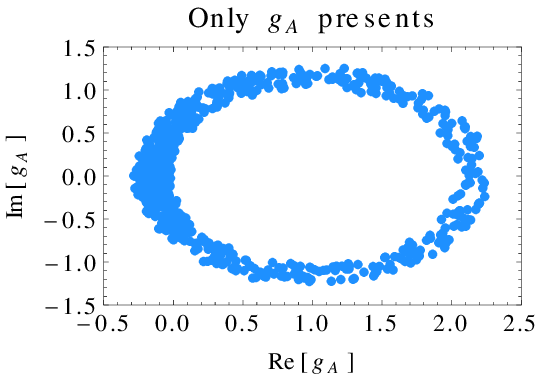}\,
\caption{The constraints on the complex coupling $g_V = V_R + V_L$  (left panel) and on the complex couplings $g_A = V_R - V_L$  (right panel) at 95\% C.L. 
\label{fig:OnlygVgA}}
\end{figure}

The combination of the couplings $g_V = V_R + V_L$ appears  in both $R(D)$ and $R(D^*)$, while $g_A = V_R - V_L$  appears only in $R(D^*)$. $V_R$ and $V_L$ receive constraints from both  $R(D)$ and $ R(D^*)$. If NP is established in both $R(D)$ and $ R(D^*)$ then the case of pure $g_A$ coupling is ruled out. The constraints on the complex couplings $g_V$  and $g_A$  are shown in the colored region of Fig.~\ref{fig:OnlygVgA} (left) and (right).
We confirm from Eq.~(\ref{tran_amp}) that if the new interaction in purely left-handed then the amplitudes and all the distributions just get scaled by a common factor. Hence, instead of considering the pure $V-A$ and $V + A$ quark current cases,  
 we will consider  cases which include pure $g_V$ or pure $g_A$ complex couplings. Interestingly,
the analysis in Ref.~\cite{Datta:2012qk} indicates that the data prefers either pure vector or pure axial vector couplings. 

\begin{figure}[h!]
\centering
\includegraphics[width=5.5cm,height=5.5cm]{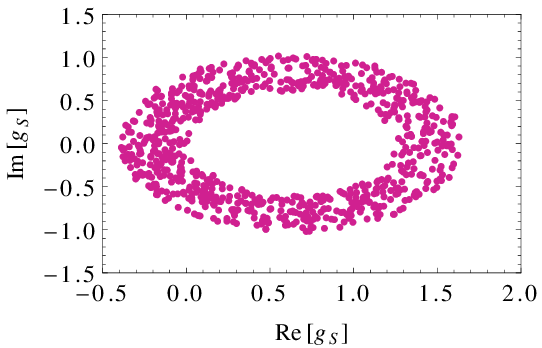}\,
\includegraphics[width=5.5cm,height=5.5cm]{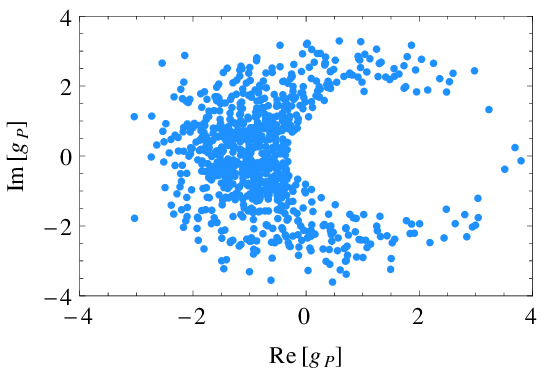}
\caption{The constraints on the  complex coupling  $g_S$ (left panel) and on the complex coupling  $g_P$ (right panel) at 95\% C.L.~. 
\label{fig:OnlygSgP}}
\end{figure}

The combination of the couplings $g_S = S_R + S_L$ appears  only  in $R(D)$, while  $ g_P = S_R - S_L$  appears only in $R(D^*)$.  If NP is established in both $R(D)$ and $ R(D^*)$ then the cases of pure
$g_S$ or $g_P$  couplings will be ruled out. 
The constraints on the complex couplings $g_S$ and $g_P$  are shown 
in Fig.~\ref{fig:OnlygSgP}.

\section{ Angular analysis}

The complete three-angle distribution for the decay  $\bar{B}\rightarrow D^{*}  (\rightarrow D \pi)l^- \bar{\nu}_l$ in the presence of NP can be expressed in terms of four kinematic variables $q^2$, two polar angles $\theta_l$, $\theta_{D^*}$, and the azimuthal angle $\chi$. The angle $ \theta_l$ is the polar angle between the charged lepton and the direction opposite to the $D^*$ meson in the $(l \nu_l)$ rest frame. The angle $\theta_{D^*}$  is the polar angle between the D meson and the direction of the $D^*$ meson in the  $(D \pi)$ rest frame. The angle $ \chi$  is the azimuthal angle between the two decay planes spanned by the 3-momenta of the $(D \pi)$ and $(l \nu_l)$ systems. These angles are described in Fig.~\ref{fig-DstlnuAD}. The three-angle distribution can be obtained by using the helicity formalism.


\noindent We can write the angular distribution explicitly for easy comparison with previous literature \cite{Dungel:2010uk, Richman:1995wm,Korner:1989qb,Korner:1989qa}
\bea
\label{3-foldAD_ex}
\frac{d^4\Gamma}{dq^2\, d\cos\theta_l\, d\cos\theta_{D^*}\, d\chi} & = &
 \frac{9}{32 \pi} NF   \Big(\sum^8_{i = 1} I_i + \frac{m_l^2}{q^2} \sum^8_{j = 1} J_i \Big), \nl
\eea
where
\bea
I_1 &= & 4\cVsq \slsq |\A0|^2, \nl
J_1 & = & 4\cVsq\Big[ |\A0|^2 \clsq + |\AtP|^2-2Re[\AtP\A0^*]\cl \Big], \nl
I_2 & = & \sVsq \Big[ (|\Apar|^2 +|\Aperp|^2 )(1+\clsq)- 4Re[\Apar\Aperp^*]\cl \Big],\nl
J_2 & = & \sVsq \slsq (|\Apar|^2 +|\Aperp|^2 ), \nl
I_3 & = & -\sVsq \slsq \cos{2 \chi} (|\Apar|^2 -|\Aperp|^2 ),\nl
J_3 & = & \sVsq \slsq \cos{2 \chi} (|\Apar|^2 -|\Aperp|^2 ),\nl 
I_4 & = & -2\sqrt{2} \ssV \sl \cos{\chi}  Re[\Aperp \A0^*],  \nl
J_4 & = & 2\sqrt{2} \ssV \sl \cos{\chi}  Re[\Apar \AtP^*],  \nl
I_5 & = & 2\sqrt{2} \ssV \sl \cl \cos{\chi} Re[\Apar \A0^*],  \nl
J_5 & = & -2\sqrt{2} \ssV \sl \cl \cos{\chi}  Re[\Apar \A0^*],  \nl
I_6 & = & \sin^2\thD \sin^2\thl \sin 2\chi Im[\Apar\Aperp^*],\nl
J_6 & = & -\sin^2\thD \sin^2\thl \sin 2\chi Im[\Apar\Aperp^*],\nl
I_7 & = &   -2 \sqrt{2}\sin 2\thD \sin\thl \sin\chi Im[\Apar\A0^*],\nl
J_7 & = &  -2 \sqrt{2}\sin 2\thD \sin\thl \sin\chi Im[\Aperp\AtP^*], \nl
I_8 & = & \sqrt{2}  \sin 2\thD \sin 2\thl \sin\chi Im[\Aperp\A0^*], \nl
J_8 & = & -\sqrt{2}  \sin 2\thD \sin 2\thl \sin\chi Im[\Aperp\A0^*],\
\eea
where the quantity $N_F$ is
\bea
\label{NF}
N_F &=& \Big[ \frac{G^2_F |p_{D^*}| |V_{cb}|^2 q^2 }{3\times 2^{6}\pi^3 m^2_B} \Big(1-\frac{m_l^2}{q^2}\Big)^2 ~Br(D^*\rightarrow D\pi)\Big] \; .
\eea
The momentum of the $D^{*}$ meson in the B meson rest frame is denoted as $|p_{D^{*}}|=\lambda^{1/2}(m^2_B,m^2_{D^{*}},q^2)/2 m_B$  with  $\lambda(a,b,c) = a^2 + b^2 + c^2 - 2 (ab + bc + ca)$. 
When there are no strong phases then $\Apar$ and $\A0$ have the same weak phase and
$I_7$ vanishes.

\FIGURE[h!]{
\centerline{
\includegraphics[width=9.5cm]{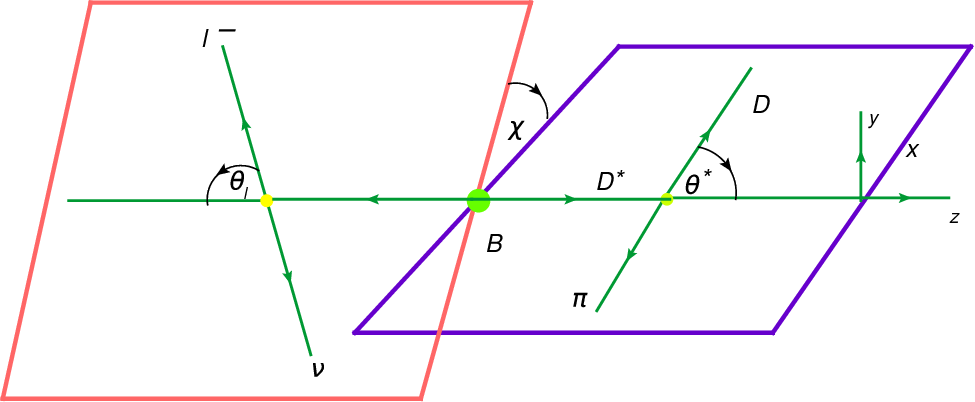}}
      \caption{The description of the angles $\theta_{\l,D^{*}}$ and $\chi$ in the angular distribution of $\bar{B}\rightarrow D^{*}  (\rightarrow D \pi)l^-\nu_l$  decay.}
\label{fig-DstlnuAD}
}

The complex NP couplings lead to CP violation which are sensitive to the angular terms
$\sin { \chi}$ and $\sin{2 \chi}$. The coefficients of these terms are 
TPs  and have the structure
$\sim Im[{\cal{A}}_i{\cal{A}}_j^*] \sim \sin( \phi_i-\phi_j) $, where 
${\cal{A}}_{i,j} =|{\cal{A}}_{i,j}|e^{ i \phi_{i,j}} $. In the SM these terms vanish, to a very good approximation, as there is only one dominant contribution to the decay and so all amplitudes have the same weak phase. 
Hence any non-zero measurements of the TPs are clear signs of NP without any hadronic uncertainties.
For the charged conjugate modes, the weak phases change sign
and ${\cal{\bar{A}}}_{i,j} =|{\cal{A}}_{i,j}|e^{- i \phi_{i,j}} $ and  the TPs change sign. Even though we focus on $\tau$ final states, we should point out that this distribution is applicable also for $e$ and $\mu$ in the final state.
Since experiments have already studied this distribution for $e$, $\mu$ final states
it might be worth checking the $\sin { \chi}$ and $\sin{2 \chi}$ terms in the distributions for these decays for signals of non-SM physics.

It will be convenient to rewrite the angular distribution as \cite{Alok:2011gv},
\begin{align}
\label{3-foldAD}
\frac{d^4\Gamma}{dq^2\, d\cos\theta_l\, d\cos\theta_{D^*}\, d\chi}=
 \frac{9}{32 \pi} NF  & \Bigg\lbrace \cos^2\thD \Big(V^0_{1} + V^0_{2}\cos 2\thl + 
     V^{0}_{3} \cos\thl \Big) \nl & + 
 \sin^2\thD \Big(V^{T}_{1} + V^{T}_{2} \cos 2\thl + 
 V^{T}_{3}\cos\thl \Big) \nl & + 
   V^{T}_{4} \sin^2\thD \sin^2\thl  \cos 2\chi + 
   V_1^{0T} \sin 2\thD \sin 2\thl   \cos\chi  \nl &+ 
   V_2^{0T}  \sin 2\thD \sin\thl  \cos\chi + 
   V^{T}_{5} \sin^2\thD \sin^2\thl \sin 2\chi \nl &+ 
   V_3^{0T}  \sin 2\thD \sin\thl \sin\chi + 
   V_4^{0T}  \sin 2\thD \sin 2\thl \sin\chi\Bigg\rbrace \; .
   \end{align}
The decay $\bar{B}\rightarrow D^{*}  (\rightarrow D \pi)l^- \bar{\nu}_l$  is completely described in terms
of twelve angular coefficient functions $V_i$. These angular coefficients depend on the couplings, kinematic variables and form factors, and are given in  the Appendix in Eq.~(\ref{eq:V0}), Eq.~(\ref{eq:VT}) and Eq.~(\ref{eq:VLT}). 

For the CP-conjugate decay $B \rightarrow \bar{D}^{*}  (\rightarrow D \pi)l^+ \nu_l$, one defines the angles relative to the directions
of the $\tau^+$ and $\bar{D}^*$.  The $\bar{V}_i$'s can be obtained from the $V_i$'s by replacing $\theta_l \to \theta_l + \pi$ and $\chi \to -\chi$, and changing the signs of the weak phases. This transformation is equivalent to replacing $V^{0}_{1,2} \to \bar{V}^{0}_{1,2}$, $V^{0}_{3} \to -\bar{V}^{0}_{3}$, $V^{T}_{1,2,4} \to \bar{V}^{T}_{1,2,4}$, $V^{T}_{3,5} \to -\bar{V}^{T}_{3,5}$, $V^{0T}_{1,3} \to \bar{V}^{0T}_{1,3}$, and $V^{0T}_{2,4} \to -\bar{V}^{0T}_{2,4}$. The angular distribution for the CP-conjugate process is 
\begin{align}
\label{3-foldAbarD}
\frac{d^4\bar{\Gamma}}{dq^2\, d\cos\theta_l\, d\cos\theta_{D^*}\, d\chi}=
 \frac{9}{32 \pi} NF  & \Bigg\lbrace \cos^2\thD \Big(\bar{V}^0_{1} + \bar{V}^0_{2}\cos 2\thl -
    \bar{V}^{0}_{3} \cos\thl \Big) \nl & + 
 \sin^2\thD \Big(\bar{V}^{T}_{1} + \bar{V}^{T}_{2} \cos 2\thl -
 \bar{V}^{T}_{3}\cos\thl \Big) \nl & + 
  \bar{V}^{T}_{4} \sin^2\thD \sin^2\thl  \cos 2\chi + 
   \bar{V}_1^{0T} \sin 2\thD \sin 2\thl   \cos\chi  \nl &- 
  \bar{V}_2^{0T}  \sin 2\thD \sin\thl  \cos\chi -
   \bar{V}^{T}_{5} \sin^2\thD \sin^2\thl \sin 2\chi \nl &+ 
  \bar{V}_3^{0T}  \sin 2\thD \sin\thl \sin\chi -
  \bar{V}_4^{0T}  \sin 2\thD \sin 2\thl \sin\chi\Bigg\rbrace \; .
   \end{align}

 \subsection{Differential branching ratio }
The angular distribution allows us to define several observables. The starting point is to obtain the differential distribution  $d\Gamma/dq^2$  after performing integration over all the angles 
\bea
\label{eq1:DBRDstsq}
\frac{d\Gamma}{dq^2 } &=& \frac{3 N_F}{4} 
(A_{L}+A_{T}) \,,\nl
\eea
where the $D^*$ longitudinal and transverse polarization amplitudes $A_L$ and $A_T$ are
\bea
\label{HL}
A_L &=& \Big(V_1^{0}  - \frac{1}{3} V_2^{0}  \Big),\quad A_T = 2 \Big(V_1^{T}  - \frac{1}{3} V_2^{T} \Big) ~.
\eea
One can see from Eq.~(\ref{eq:V0}) and Eq.~(\ref{eq:VT}) that $A_L$ is proportional to $|\A0|^2$ and $|\AtP|^2$ while $A_T$ to $|\Apar|^2 + |\Aperp|^2 $.  The $D^*$ polarization amplitudes can be extracted
from the angular distribution in $\cV$ (see Eq.~(\ref{cDstAD}) below). Since there is no direct CP violation,
we have $A_{L,T} = \bar{A}_{L,T}$.
Hence,
\bea
\label{eq:RDst}
 ~~ \frac{d\Gamma}{dq^2 } &= & \frac{d \bar{\Gamma}}{dq^2 } \,.
\eea
Furthermore, one can also explore the $q^2$ dependent ratio
\bea
\label{eq20:DDRRDst2}
R_{D^*}(q^2) &=&\frac{d Br[\BDstartaunu]/dq^2 }{d Br[\BDstarlnu]/dq^2}\,,
\eea
where $l$ denotes the light lepton $(e, \mu)$. The ratio  $R_{D^*}$  are independent of the  form factor $h_{A_1}(w)$. 
 \FIGURE[t]{
\includegraphics[width=0.4\linewidth]{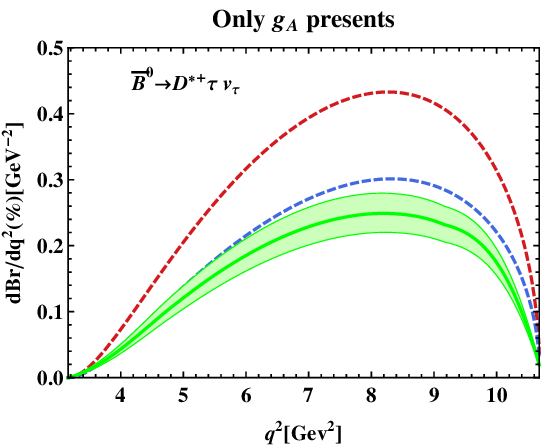}
\includegraphics[width=0.4\linewidth]{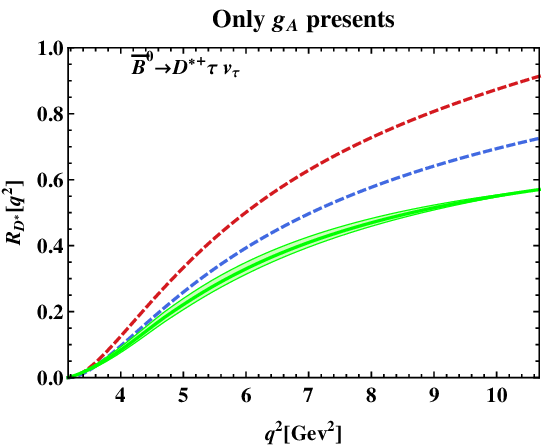} \\
  \includegraphics[width=0.4\linewidth]{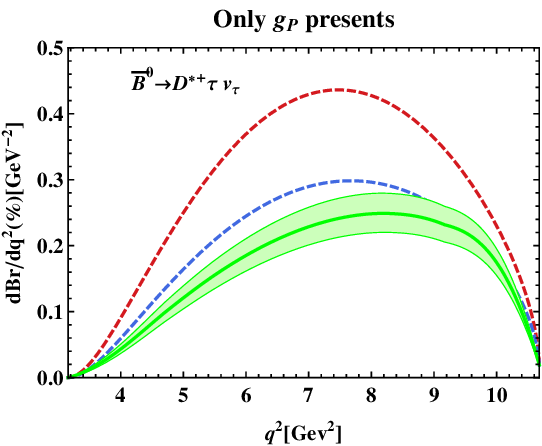}
 \includegraphics[width=0.4\linewidth]{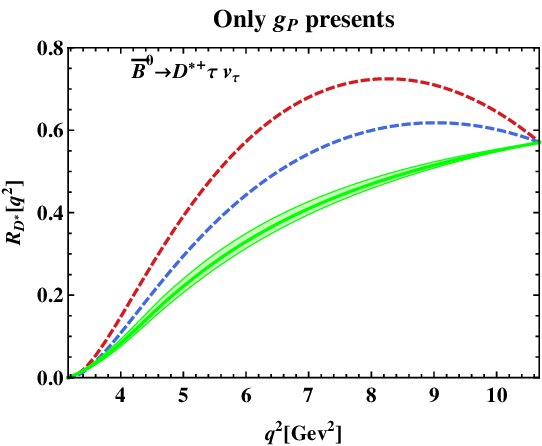}
\caption{The left (right) panels of the figure show the differential branching ratio ($R_{D^*}(q^2)$)  for the decay $\bar{B}^0 \to D^{*+} \tau \nu_\tau$ in the scenario   where only $g_A$ (upper) and only $g_P$ (lower) couplings  are present.  The
green  band corresponds to the SM prediction and its uncertainties.   The red  and blue dashed lines correspond to $|g_A| e^{i \phi_{g_A}} = 0.62 e^{i 1.74} $ and $|g_A| e^{i \phi_{g_A}} = 0.13 e^{i 2.63} $ respectively, in the upper-left panel, and $|g_A| e^{i \phi_{g_A}} = 0.34 e^{i 2.38} $ and $|g_A| e^{i \phi_{g_A}} = 0.73 e^{i 1.39} $ in the upper-right panel. The red  and blue dashed lines correspond to $|g_P| e^{i \phi_{g_P}} = 2.27 e^{-i 2.92} $ and $|g_P| e^{i \phi_{g_P}} = 1.65 e^{-i 2.96} $ in the lower-left panel, and $|g_P| e^{i \phi_{g_P}} = 2.56 e^{-i 2.17} $ and $|g_P| e^{i \phi_{g_P}} = 2.60 e^{-i 1.95} $ in the lower-right panel. The values of the couplings are chosen to show the maximum and minimum deviations from the SM expectations.   
\label{fig:dBRgAgP}}
}

Fig.~\ref{fig:dBRgAgP} shows the differential branching ratio (DBR)  and $R_{D^*}(q^2)$ for $\bar{B}^0 \to D^{*+} \tau \nu_\tau$ in the presence of only $g_A = V_R -V_L$ and only $g_P = S_R -S_L$ couplings. We make the following observations:
\begin{itemize}

\item If only the $g_A$ coupling is present, the DBR can be  enhanced up to 0.4\% at  $q^2 \approx 8.5 \mathrm{GeV}^2$. $R_{D^*}(q^2)$ can be enhanced up to 0.9\% at high $q^2$. The shape of the distribution is similar to that in the SM.

 \item If only the $g_P$ coupling is present, the DBR can be  enhanced up to 0.4\% at  $q^2 \approx 7.5 \mathrm{GeV}^2$. Note that the peak of the DBR is shifted to the low $q^2$ direction relative to the SM.  $R_{D^*}(q^2)$  is approximately 0.7  at  $q^2 \approx 7.5 \mathrm{GeV}^2$. The shape of the distribution is different from that in the SM.
\end{itemize}

Finally,  the new NP coupling $g_V$ only appears in the transverse amplitude ${\cal{A}}_\perp$, and does not significantly affect the DBR  and $R_{D^*}(q^2)$.
The shape of the distribution is again similar to that in the SM.

We note that recently BaBar has reported  the measurement of the differential distribution  for both $\BDstartaunu$ and $\BDtaunu$ decays \cite{Lees:2013uzd} and the results seem to generally favor vector, axial-vector type NP though scalar/pseudoscalar NP are not ruled out.

\subsection{Polarization fraction for $D^*$ }

The differential angular distribution in $\cV$ gives access to the polarization fraction of the $D^{*}$ meson in the decay $\bar{B}\rightarrow D^{*}  (\rightarrow D \pi)\tau^- \bar{\nu}_\tau$ 
\bea
\label{cDstAD}
\frac{d^2 \Gamma}{dq^2 d\cos{\thD}} = 
\frac{1}{4} \frac{d\Gamma}{dq^2 } (2 F^{D^*}_L \cos^2{\thD} + (1- F^{D^*}_L) \sin^2{\thD})\,,
\eea
where we  define the longitudinal and transverse polarization fractions of the $D^*$ meson  as
\bea
\label{eq17:DDRBstfLDst}
F^{D^*}_L(q^2) &=& \frac{A_L}{A_L +A_T }\,,\quad \quad 
F^{D^*}_T(q^2) = \frac{A_T}{A_L +A_T }\,,
\eea
with $F^{D^*}_L(q^2) + F^{D^*}_T(q^2) = 1$. Similarly, one can define the polarization fractions $\bar{F}^{D^*}_{L,T}(q^2)$ for the CP-conjugate mode but they are the same as $F^{D^*}_{L,T}(q^2)$ in the absence of direct CP violation. 

\begin{figure}[h!]
\centering
\includegraphics[width=0.4\linewidth]{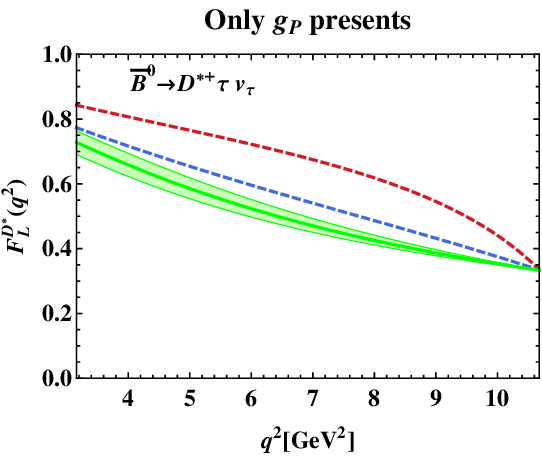}
\includegraphics[width=0.4\linewidth]{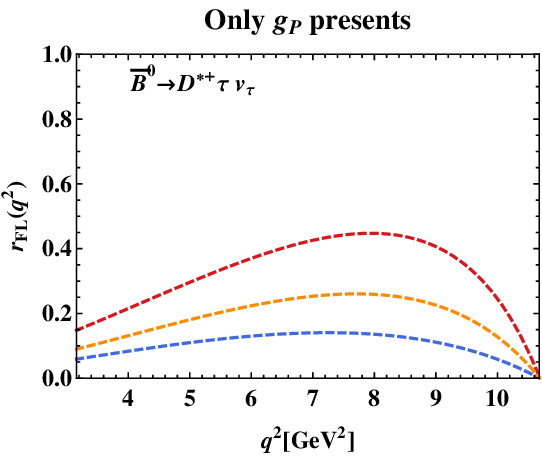}
\caption{The left  and right panels of the figure show $F_{L}^{D^*}$ and the ratio $r_{F_L}$ for the decay 
$\BDstartaunu$ in the scenario   where  only the $g_P$   coupling  is present.  The green  band corresponds to the SM prediction and its uncertainties. The red  and blue dashed lines correspond to $|g_P| e^{i \phi_{g_P}} = 2.27 e^{-i 2.92} $ and $|g_P| e^{i \phi_{g_P}} = 0.90 e^{-i 2.74} $ respectively .
The values of the couplings are chosen to show the maximum and minimum deviations from the SM expectations.  
\label{fig:FLgP}}
\end{figure}

When only the coupling $g_A$ is present, the polarization fractions  of the $D^*$ meson gets contributions from the amplitudes $\A0$ and $\Apar$, which are functions of the new $g_A$ coupling.  Due to cancellation of the $g_A$ coupling contributions in $F^{D^*}_L(q^2)$, its  behavior is similar to its SM prediction. New $g_V$ coupling appears only in the amplitude $A_{\perp}$ and again $F^{D^*}_L(q^2)$ looks similar to its SM prediction. In Fig.~\ref{fig:FLgP}, 
we show $F^{D^*}_L(q^2)$  and the ratio $r_{F_L} =[F^{D^*}_L]^{NP+SM}/[F^{D^*}_L]^{SM}- 1  = [F^{D^*}_L]^{NP}/[F^{D^*}_L]^{SM}$ in the presence of  the $g_P$ coupling. In this case $F^{D^*}_L(q^2)$ can be as large as 0.85 at low $q^2$, and it decreases to the SM value at high $q^2$ while  the ratio $r_{F_L}$  reaches  $40\%$  around  $q^2 \approx 8.0 $ $\mathrm{GeV}^2$.

The $q^2$-integrated  polarization fractions $<F^{D^*}_{L,T}>$ can be obtained by separately integrating out the numerator and the denominator in Eq.~(\ref{eq17:DDRBstfLDst}). We obtain $<F^{D^*}_{L}>_{SM} \approx 0. 53 $ in the SM for the $\bar{B}^0 \to D^{*+} \tau \nu_\tau$ decay and only the new  $g_P$ coupling can enhance $<F^{D^*}_{L}>$ by about $6\%$ from its SM value. 

\subsection{Distribution in $\cl$ and $A_{FB}$ }
The forward-backward asymmetry $A_{FB}$  can be  obtained  from  the  single-differential angular distribution

\begin{align}
\label{doubDRAFB}
\frac{d^2\Gamma}{dq^2d\cos{\theta_l}  } &=&\frac{3 N_F}{4} \Big[\Big(V^{T}_1 + \frac{1}{2}  V^{0}_1) + \Big(V^{T}_2 + \frac{1}{2}  V^{0}_2 \Big) \cos{2  \theta_l} + \Big( 
      V^{T}_3 + \frac{1}{2} V^{0}_3)  \cos{\theta_l} \Big]\,,
 \end{align}

The forward-backward asymmetry (FBA) for the leptons is defined by
\bea
\label{FBA}
A_{FB}(q^2) &=&\frac{\int^1_0 d\cos{\theta_l}\frac{d^2\Gamma}{dq^2d\cos{\theta_l}  }-\int^0_{-1} d\cos{\theta_l}\frac{d^2\Gamma}{dq^2d\cos{\theta_l}  }}{\int^1_0 d\cos{\theta_l} \frac{d^2\Gamma}{dq^2d\cos{\theta_l}  }+\int^0_{-1} d\cos{\theta_l}\frac{d^2\Gamma}{dq^2d\cos{\theta_l}  }} \; .
\eea
Then one can obtain:
\bea
\label{FBA1}
A_{FB}(q^2) = \frac{ V^{T}_3 + \frac{1}{2} V^{0}_3}{A_L + A_T}\,,
\eea
similarly, FBA for the conjugate mode is
\bea
\label{FBA2}
\bar{A}_{FB}(q^2) = -\Big[\frac{ \bar{V}^{T}_3 + \frac{1}{2} \bar{V}^{0}_3}{\bar{A}_L + \bar{A}_T}\Big]\,.
\eea
If the absence of direct CP violation, $\bar{A}_{FB}(q^2) = -{A}_{FB}(q^2) $.
 We define the average FBA as 
\bea
\label{FBA3}
A_{FB}^{D^*}(q^2) = \frac{1}{2} \Big(A_{FB}(q^2) - \bar{A}_{FB}(q^2) \Big)\,.
\eea
 
Within the SM, $A_{FB}^{D^*}(q^2)$ has a zero crossing at $q^2 \approx 5.64 \mathrm{GeV}^2$ (see Fig.~(\ref{fig:AFBgAgVgP})). We make the following observations from this figure
\begin{itemize}

\item If only the $g_A$ or only the $g_V$ couplings are present, the FBA can reach a value close to 50\% at low  $q^2$ and its sign is mostly negative. The FBA converges  to its SM prediction  at high $q^2$.

 \item If only the $g_P$ coupling is present, the FBA can reach a value up to 30\% at  low $q^2$. It can have both positive or negative signs. Again, the FBA converges  to its SM prediction  at high $q^2$.
\end{itemize}
 \FIGURE[t]{
\includegraphics[width=0.4\linewidth]{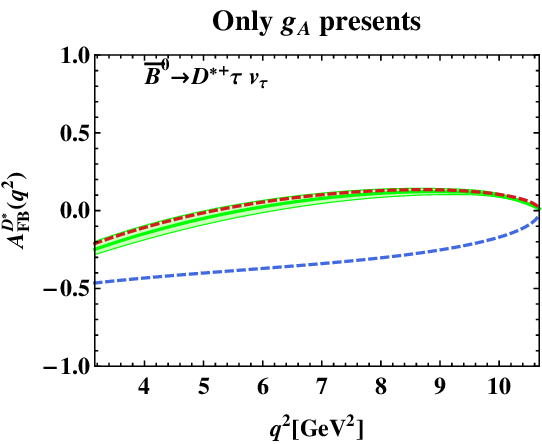}
\includegraphics[width=0.4\linewidth]{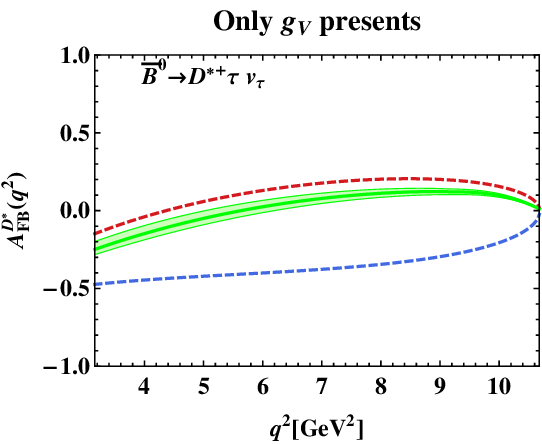} \\
  \includegraphics[width=0.4\linewidth]{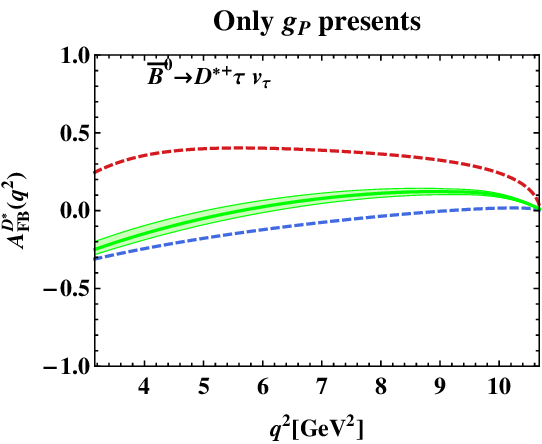}
 \caption{The figures show   $A_{FB}^{D^*}(q^2)$ for the decay
$\BDstartaunu$ in the scenario   where only $g_A$,  $g_V$ and  $g_P$  couplings  are present.  The
green  band corresponds to the SM prediction and its uncertainties. The red  and blue dashed lines correspond to $|g_A| e^{i \phi_{g_A}} = 0.1 e^{i 2.19} $ and $|g_A| e^{i \phi_{g_A}} = 2.06 e^{i 0.1} $ respectively, in the upper-left panel,  $|g_V| e^{i \phi_{g_V}} = 0.48 e^{-i 0.84} $ and $|g_V| e^{i \phi_{g_V}} = 2.23 e^{i 2.94} $ in the upper-right panel, and  $|g_P| e^{i \phi_{g_P}} = 3.53 e^{-i 0.11} $ and $|g_P| e^{i \phi_{g_P}} = 1.69 e^{-i 2.94} $  in the lower panel. The values of the couplings are chosen to show the maximum and minimum deviations from the SM expectations.  
\label{fig:AFBgAgVgP}}
}

In Table.\ref{tabAFBint} we summarize the predictions for the $q^2$-integrated FBA $ \langle A_{FB}^{D^*} \rangle $  for the decay  $\bar{B}^0 \to D^{*+} \tau \nu_\tau$.
\begin{table}[t]
\begin{center}
\caption{\label{tabAFBint} \it \small Predictions for the $q^2$-integrated FBA $ \langle A_{FB}^{D^*} \rangle $ both within the SM and in the presence of different NP couplings for the decay  $\bar{B}^0 \to D^{*+} \tau \nu_\tau$.}
\vspace{0.2cm}
\doublerulesep 0.8pt \tabcolsep 0.08in
\small{
\begin{tabular}{l c c c c}\hline\hline
	        &   SM Prediction                       &   Only $g_A$	
                    &   Only $g_V$	                        &   Only $g_P$
\\ \hline\\[-10pt]
$ \langle A_{FB}^{D^*} \rangle $ 	        & $-0.041$           & $[-0.055, -0.349]$
                    & $[-0.382,0.045]$	    & $[-0.127,0.343]$\\
\\
\hline\hline
\end{tabular}
}
\end{center}
\end{table}

\subsection{\boldmath Asymmetries $A_C^{(i)}$ in the angular variable $\chi$  }
In this section, we consider three different transverse asymmetries $A_C^{(i=1,2,3)}$. These asymmetries are obtained by integrating out the polar angles $\thl$ and $\thD$ in  three different regions.

\subsubsection{$A_C^{(1)}$}
\begin{figure}[h!]
\centering
\includegraphics[width=0.4\linewidth]{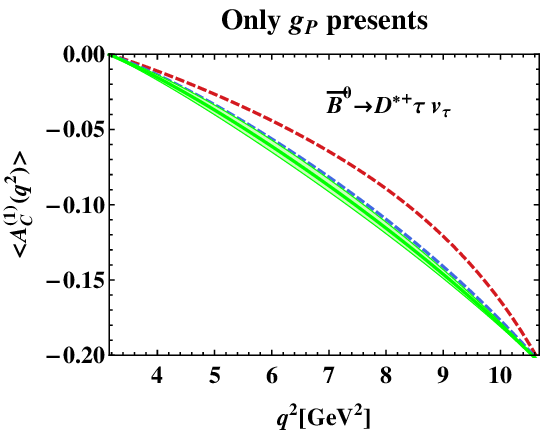}
\includegraphics[width=0.4\linewidth]{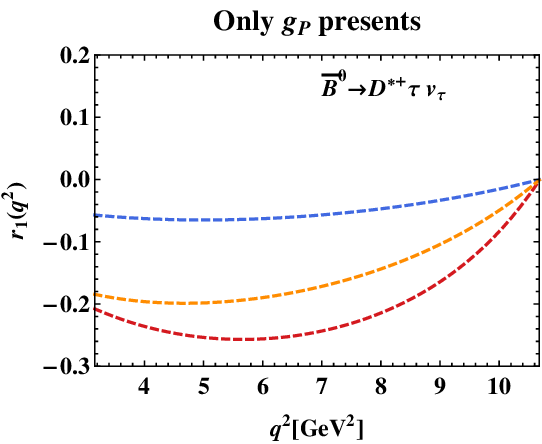}
\caption{The figure shows   $\langle A^{(1)}_C(q^2)\rangle$ and $r_1(q^2)$ for the decay
$\BDstartaunu$  in the scenario   where  only the $g_P$   coupling  is present.  The green  band corresponds to the SM prediction and its uncertainties. The red  and blue dashed lines correspond to $|g_P| e^{i \phi_{g_P}} = 2.17 e^{i 1.75} $ and $|g_P| e^{i \phi_{g_P}} = 0.68 e^{-i 1.79} $ respectively . The values of the couplings are chosen to show the maximum and minimum deviations from the SM expectations.    
\label{fig:AC1gP}}
\end{figure}

The transverse asymmetry $A_C^{(1)}$ is defined through the angular distribution in $\chi$ as
\bea
\label{doubDRAC1}
\frac{d^2\Gamma}{dq^2 d\chi } &=&\frac{1}{2\pi} \frac{d\Gamma}{dq^2  }
\Big[ 1+  \left(A^{(1)}_C \cos{2 \chi} + A^{(1)}_T \sin{2 \chi}\right)
\Big] \; .
\eea
It can be obtained by integrating Eq.~(\ref{3-foldAD}) over the two polar
angles $\thl$ and $\thD$. Here $A^{(1)}_T$  is
a TP, and is discussed separately below.  One can obtain
\bea
 \label{eq1:AC1}
A^{(1)}_C (q^2)&=& \frac{4 V^T_4}{3 (A_L+ A_T)} \,,
\eea
 and similarly,  for the conjugate mode 
\bea
\label{eq2:AC1}
\bar{A}^{(1)}_C (q^2)&=& \frac{4 \bar{V}^T_4}{3 (\bar{A}_L + \bar{A}_T)} \,.
\eea
In the absence of direct CP violation $\bar{A}^{(1)}_C  ={A}^{(1)}_C $. 
 We define the average $A^{(1)}_C (q^2)$ as
\bea
\label{eq3:AC1}
\langle A^{(1)}_C(q^2) \rangle  = \frac{1}{2} \Big(A^{(1)}_C(q^2) + \bar{A}^{(1)}_C (q^2) \Big)\,.
\eea
The SM prediction for $\langle A^{(1)}_C(q^2)\rangle$ is shown in the green band of Fig.~\ref{fig:AC1gP}(left panel). Similar to  $F^{D^*}_{L}$, the asymmetry $\langle A^{(1)}_C(q^2)\rangle$ remains almost the same as the SM prediction when only the $g_A$ or only the $g_V$ couplings are presents.

The   $g_P$ coupling appears only in the amplitude $A_{P}$ and affects only the denominator of $A^{(1)}_C(q^2)$.  
The amplitude ${\cal{A}}_{P}$ vanishes at high $q^2$, and hence $\langle A^{(1)}_C(q^2)\rangle$ reduces to its SM value as shown in  Fig.~\ref{fig:AC1gP}(left panel).  The magnitude of the ratio $r_1 (q^2)= [A^{(1)}_C(q^2)]^{SM+NP}/[A^{(1)}_C(q^2)]^{SM}-1= [A^{(1)}_C(q^2)]^{NP}/[A^{(1)}_C(q^2)]^{SM}$ 
reaches more than $25\%$ at $ q^2 \approx 5.0 \mathrm{GeV}^2$ 
as shown in Fig.~\ref{fig:AC1gP}(right panel).

\subsubsection{$A_C^{(2)}$}
We define the  angular distribution 

\bea
\label{eq1:Gam2}
\frac{d^2 \Gamma^{(2)}}{dq^2 d\chi} & = & \Big[\int^1_0 - \int^{0}_{-1} \Big] \frac{d^4\Gamma}{dq^2\, d\cos\theta_l\, d\cos\theta_{D^*}\, d\chi} ~d\cos\theta_{D^*}\,.
\eea

\begin{figure}[h!]
\centering
\includegraphics[width=0.4\linewidth]{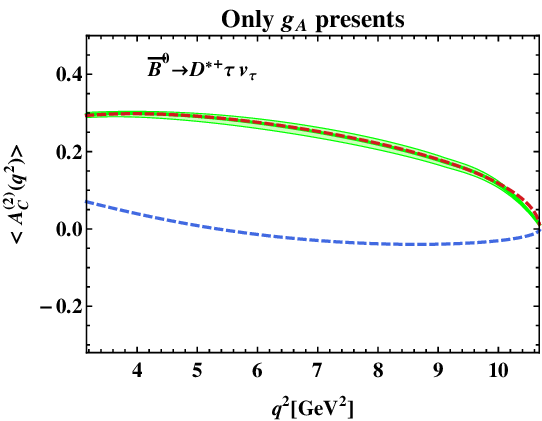}
\includegraphics[width=0.4\linewidth]{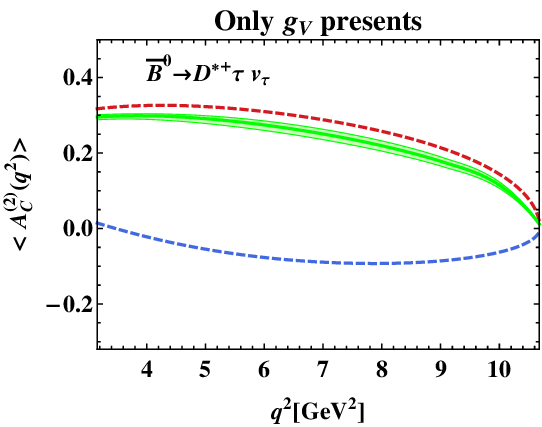} \\
  \includegraphics[width=0.4\linewidth]{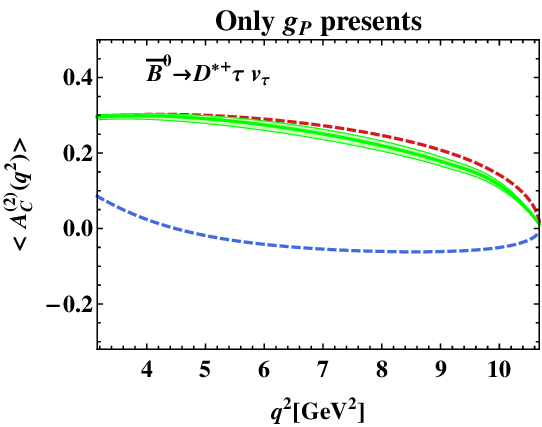}
 \caption{The figures show  $\langle A^{(2)}_C (q^2)\rangle$ for the decay $\bar{B}^0 \to D^{*+} \tau \nu_\tau$ in the scenario   where only $g_A$ , only $g_V$ and only $g_P$  couplings  are present.  The green  band corresponds to the SM prediction and its uncertainties. The red  and blue dashed lines correspond to $|g_A| e^{i \phi_{g_A}} = 0.04 e^{i 2.83} $ and $|g_A| e^{i \phi_{g_A}} = 2.06 e^{i 0.1} $ respectively, in the upper-left panel,  $|g_V| e^{i \phi_{g_V}} = 0.34 e^{i 0.28} $ and $|g_V| e^{i \phi_{g_V}} = 2.37 e^{-i 3.12} $ in the upper-right panel, and  $|g_P| e^{i \phi_{g_P}} = 0.82 e^{-i 2.67} $ and $|g_P| e^{i \phi_{g_P}} = 3.80 e^{-i 0.04} $  in the lower panel. The values of the couplings are chosen to show the maximum and minimum deviations from the SM expectations.  
\label{fig:AC2gAgVgP}}
\end{figure}

One can obtain
\bea
\label{eq2:Gam2}
\frac{d^2 \Gamma^{(2)}}{dq^2 d\chi} & = &
\frac{1}{4} \frac{d\Gamma}{dq^2} \Big[ A_C^{(2)} \cos{\chi} + A_T^{(2)} \sin{\chi}\Big]\,,
\eea
where
\bea
\label{eq1:AC2}
A^{(2)}_C (q^2)&=& \frac{ V^{0T}_2}{ (A_L+ A_T)} \,.
\eea
Here $A^{(2)}_T$  is a TP, and is discussed separately below. For the conjugate mode 
\bea
\label{eq2:AC2}
\bar{A}^{(2)}_C (q^2)&=& \frac{- \bar{V}^{0T}_2}{ (\bar{A}_L + \bar{A}_T)} \,.
\eea
In the absence of direct CP violation $\bar{A}^{(2)}_C  = -{A}^{(2)}_C $. 
 We define the average $A^{(2)}_C (q^2)$ as 
\bea
\label{eq3:AC2}
\langle A^{(2)}_C (q^2)\rangle = \frac{1}{2} \Big(A^{(2)}_C(q^2) -\bar{A}^{(2)}_C (q^2) \Big)\,.
\eea
$\langle A^{(2)}_C (q^2)\rangle$ depends on all the three  couplings $g_A$, $g_V$, and $g_P$. For all $q^2$, the magnitude of $\langle A^{(2)}_C (q^2)\rangle$ is generally suppressed by  these new couplings. As shown in Fig.~\ref{fig:AC2gAgVgP}, in all three cases, the value of $\langle A^{(2)}_C (q^2)\rangle$  can be either  positive or negative. In particular, there may or may not be a non-SM zero crossing.

\begin{figure}[h!]
\centering
\includegraphics[width=0.4\linewidth]{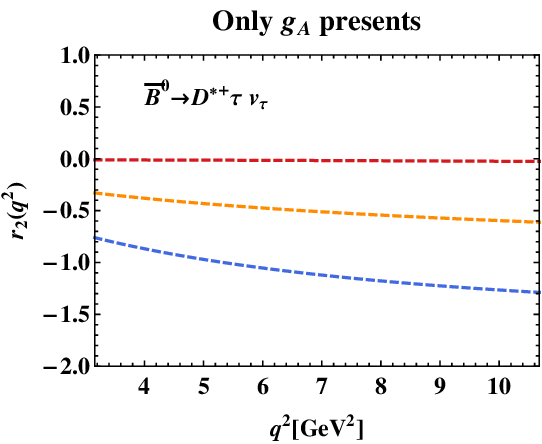}
\includegraphics[width=0.4\linewidth]{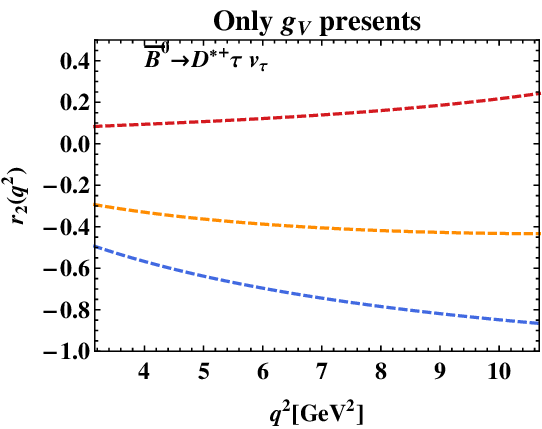} \\
  \includegraphics[width=0.4\linewidth]{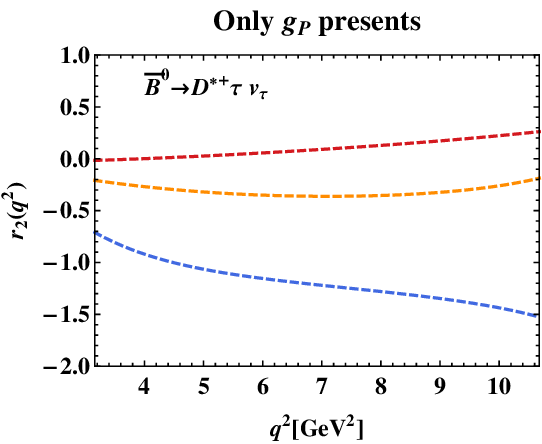}
 \caption{The figures show  $r_2 (q^2)$ for the decay $\bar{B}^0 \to D^{*+} \tau \nu_\tau$ in the scenario   where only the $g_A$ , only the $g_V$ and only the $g_P$  couplings  are present.  The values of red  and blue dashed lines  are given in the figure caption of  Fig.8.}
\label{fig:r2gAgVgP}
\end{figure}

The $q^2$ dependence  of the ratio 
$r_2 (q^2)= [A^{(2)}_C(q^2)]^{SM+NP}/ [A^{(2)}_C(q^2)]^{SM} -1 =
[A^{(2)}_C(q^2)]^{NP}/ [A^{(2)}_C(q^2)]^{SM}
$ is shown in  Fig.~\ref{fig:r2gAgVgP} for all the three cases. The magnitude of $r_2 (q^2)$ can be more than $100\%$  at high $q^2$.

\subsubsection{$A_C^{(3)}$}
Finally, we  define the single angle distribution 

\begin{align}
\label{eq1:Gam3}
\frac{d^2 \Gamma^{(3)}}{dq^2 d\chi} & = & \Big[\int^1_0 - \int^{0}_{-1} \Big]~d\cos\theta_l~\Big[\int^1_0 - \int^{0}_{-1} \Big]  d\cos\theta_{D^*}~ \frac{d^4\Gamma}{dq^2\, d\cos\theta_l\, d\cos\theta_{D^*}\, d\chi}\,.
\end{align}

One can obtain
\bea
\label{eq2:Gam3}
\frac{d^2 \Gamma^{(3)}}{dq^2 d\chi} & = &
\frac{2}{3 \pi} \frac{d\Gamma}{dq^2} \Big[ A_C^{(3)} \cos{\chi} + A_T^{(3)} \sin{\chi}\Big]\,,
\eea

where
\bea
\label{eq1:AC3}
A^{(3)}_C (q^2)&=& \frac{ V^{0T}_1}{ (A_L+ A_T)} \,.
\eea
Here $A^{(3)}_T$  is a TP, and is discussed separately below. For the conjugate mode 
\bea
\label{eq2:AC3}
\bar{A}^{(3)}_C (q^2)&=& \frac{ \bar{V}^{0T}_1}{ (\bar{A}_L + \bar{A}_T)} \,.
\eea
In the absence of direct CP violation $\bar{A}^{(3)}_C  = {A}^{(3)}_C $. 
 We define the average $A^{(3)}_C (q^2)$ as 
\bea
\label{eq3:AC3}
\langle A^{(3)}_C(q^2)\rangle = \frac{1}{2} \Big(A^{(3)}_C(q^2) + \bar{A}^{(3)}_C (q^2) \Big)\,.
\eea

\begin{figure}[h!]
\centering
\includegraphics[width=0.4\linewidth]{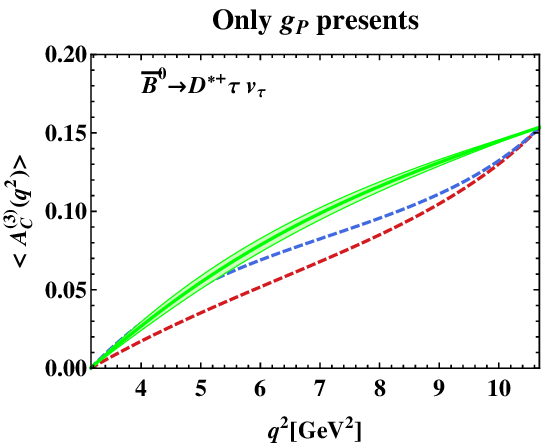}
\includegraphics[width=0.4\linewidth]{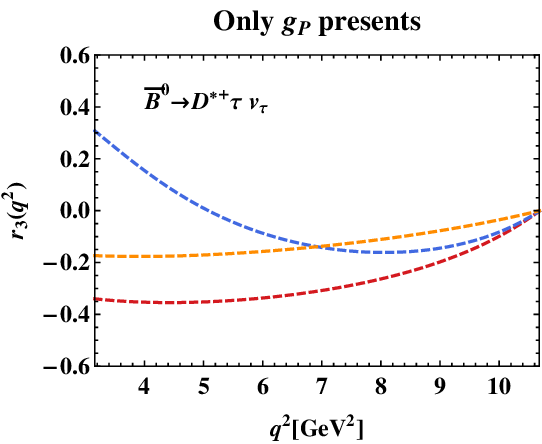}
\caption{The figure shows    $\langle A^{(3)}_C(q^2)\rangle$  and $r_3(q^2)$ for the decay $\bar{B}^0 \to D^{*+} \tau \nu_\tau$ in the scenario   where  only the $g_P$   coupling  is present.  The green  band corresponds to the SM prediction and its uncertainties. The red  and blue dashed lines correspond to $|g_P| e^{i \phi_{g_P}} = 2.03 e^{i 2.67} $ and $|g_P| e^{i \phi_{g_P}} = 3.08 e^{i 0.63} $, respectively . The values of the couplings are chosen to show the maximum and minimum deviations from the SM expectations.  
\label{fig:AC3gP}}
\end{figure}

The angular coefficient $V^{0T}_1$ depends only on the $g_A$ coupling. Due to cancellations of the NP contributions, $\langle A^{(3)}_C(q^2)\rangle$ behaves similar to its SM prediction  when only the $g_A$ coupling is present. 
The SM prediction of $\langle A^{(3)}_C(q^2)\rangle$ is shown in the  green band of  Fig.~\ref{fig:AC3gP}(left panel). $\langle A^{(3)}_C(q^2)\rangle$ is not sensitive  to the new  $g_V$ coupling. As shown in    Fig.~\ref{fig:AC3gP}(left panel),  $\langle A^{(3)}_C(q^2)\rangle$ is suppressed relative to the SM by the new $g_P$ coupling.   The magnitude of the ratio $r_3 (q^2)= [A^{(3)}_C(q^2)]^{SM+NP}/ [A^{(3)}_C(q^2)]^{SM} -1 = [A^{(3)}_C(q^2)]^{NP}/ [A^{(3)}_C(q^2)]^{SM}$ can reach values $ \gsim 30\%$ at low $ q^2 $ as shown in Fig.~\ref{fig:AC3gP}(right panel).

\subsection{CP-violating triple-product asymmetries}

In this subsection, we consider the TPs  in the
decays $\bar{B}\rightarrow D^{*}  (\rightarrow D \pi)l^- \bar{\nu}_l$ and $B \rightarrow \bar{D}^{*}  (\rightarrow D \pi)l^+ \nu_l$. For the decaying $\bar{B}$
meson, the TP is proportional to $(\hat{n}_D \times \hat{n}_l) \cdot
\hat{n}_z$ in its rest frame, where the unit vectors are given in
terms of the momenta of the final-state particles as \cite{Alok:2011gv}
\bea
\label{unitvecdef}
\hat{n}_D &=& \frac{\hat{p}_{D} \times \hat{p}_{\pi} }
{|\hat{p}_{D}\times \hat{p}_{\pi}|},~~\hat{n}_z = 
\frac{\hat{p}_{D}+ \hat{p}_{\pi}}{|\hat{p}_{D} + \hat{p}_{\pi}|} = \{0,0,1\},
~~\hat{n}_l =\frac{\hat{p}_{l^-} \times \hat{p}_{\bar{\nu}_\tau} }
{|\hat{p}_{l^-} \times \hat{p}_{\bar{\nu}_\tau} |} \; .
\eea
The vectors $\hat{n}_D$ and $\hat{n}_l$ are perpendicular to the decay planes of the $D^*$ and the virtual vector boson.  In terms of the azimuthal angle $\chi$, one gets 
\bea
\label{unitvec}
\cos{\chi} &=& \hat{n}_D \cdot \hat{n}_l\; , \quad
\sin{\chi}= (\hat{n}_D \times \hat{n}_l) \cdot \hat{n}_z \; ,
\eea
and hence the quantities that are coefficients of $\sin \chi$ (or of
$\sin 2\chi = 2 \sin\chi \cos \chi$) are the TPs.

As noted above, while the angular distribution for the $\bar{B}$ decay
involves $\chi$, for $B$ it involves $-\chi$.  The TPs in the SM vanish to a very good approximation, as we have mentioned earlier, and this result is free from any hadronic uncertainties. However,
with NP the TPs are not zero in general for complex NP couplings. The non-zero TPs now depend on the form factors and suffer from the hadronic uncertainties coming from the form factors. In our calculation for the TPs we have used the inputs for the form factors at their central values. The hadronic uncertainties in the TPs predictions are included in the range of the various NP couplings.

\subsubsection{$A_T^{(1)}$}
The first TP is $A^{(1)}_T$, introduced above in Eq.~(\ref{doubDRAC1}). One can find $A^{(1)}_T$ and $\bar{A}^{(1)}_T$ as
\bea
 \label{eq1:AT1}
 A^{(1)}_T  (q^2)&=& \frac{4 V^T_5}{3(A_L+ A_T)},\quad \bar{A}^{(1)}_T (q^2) = -\frac{4 \bar{V}^T_5}{3 (\bar{A}_L + \bar{A}_T)} \; . 
\eea
In the absence of direct CP violation $\bar{A}^{(1)}_T   = A^{(1)}_T  $. 
We observe that $A^{(1)}_T$ depends  on both the  $g_A$ and the  $g_V$ couplings
 and not on the $g_P$ coupling. The
CP-violating triple-product asymmetry is
\bea
\label{eq2:AT1}
\langle A_T^{(1)} (q^2) \rangle &=&  \frac{1}{2} \Big(A_T^{(1)} (q^2)+ \bar{A}_T^{(1)} (q^2)\Big)~.
\eea
Fig.~\ref{fig:AT1gAgV} shows $\langle A_T^{(1)} (q^2)\rangle$  for $\BDstartaunu$ in the presence of only the $g_A$ and only the  $g_V $ couplings. We make the following observations:
\begin{itemize}

\item If only the $g_A$ coupling is present, the magnitude of  $\langle A_T^{(1)}(q^2) \rangle$ can be  enhanced up to 4\% at  $q^2 \approx 8.0 \mathrm{GeV}^2$. It vanishes at the end points as the amplitude ${\cal{A}}_\perp$ diminishes.  $\langle A_T^{(1)} (q^2)\rangle$ can be either positive or negative. It may or may not have non-SM zero crossing points.

 \item If only the $g_V$ couplings is present, the magnitude of  $\langle A_T^{(1)}(q^2) \rangle$ can be  enhanced up to 5\% at  $q^2 \approx 8.0 \mathrm{GeV}^2$.  The behavior of  $\langle A_T^{(1)}(q^2) \rangle$ is similar to the above case.
\end{itemize}

\FIGURE[t]{
\includegraphics[width=0.4\linewidth]{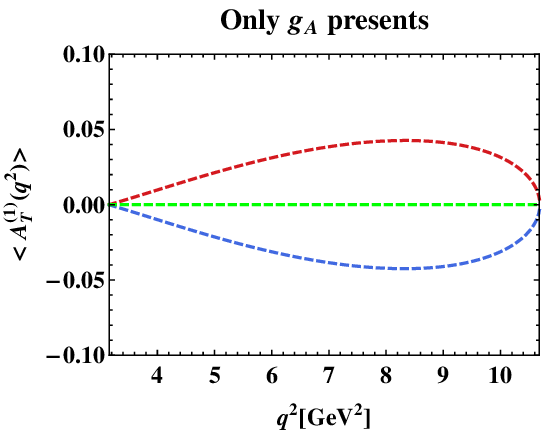}
\includegraphics[width=0.4\linewidth]{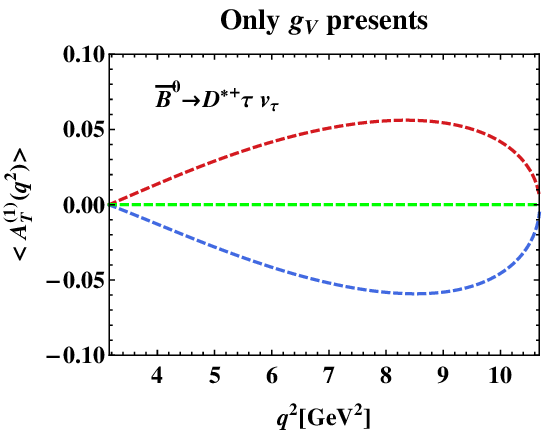}
 \caption{The left and right panels of the figure show  $\langle A_T^{(1)} (q^2)\rangle$ for the decay $\BDstartaunu$ in the scenario   where only the $g_A$  and only the $g_V$  couplings  are present. The green  dashed line corresponds to the SM prediction. The red  and blue dashed lines correspond to $|g_A| e^{i \phi_{g_A}} = 1.3 e^{i 0.92} $ and $|g_A| e^{i \phi_{g_A}} = 1.56 e^{-i 0.74} $ respectively, in the left panel, and  $|g_V| e^{i \phi_{g_V}} = 1.73 e^{i 2.34} $ and $|g_V| e^{i \phi_{g_V}} = 1.75 e^{-i 2.25} $ in the right panel. The values of the couplings are chosen to show the maximum and minimum deviations from the SM expectations.  
\label{fig:AT1gAgV}}
}

\subsubsection{$A_T^{(2)}$}
The second TP is $A_{T}^{(2)}$, introduced above in Eq.~(\ref{eq2:Gam2}). $A_{T}^{(2)}$  and $\bar{A}_{T}^{(2)}$ are given by
\bea
 \label{eq1:AT2}
 A^{(2)}_T(q^2) &=& \frac{ V^{0T}_3}{(A_L+ A_T)},\quad \bar{A}^{(2)}_T = \frac{ \bar{V}^{0T}_3}{(\bar{A}_L + \bar{A}_T)} \; . 
\eea
We observe that $A^{(2)}_T(q^2)$ depends  on all the three new couplings $g_A$, $g_V$,  
and  $g_P$. This TP is proportional to the lepton mass and so is very small when the lepton is the electron or the muon. The
CP-violating triple-product asymmetry is
\bea
\label{eq2:AT2}
\langle A_{T}^{(2)} (q^2)\rangle &=& \frac{1}{2} \Big( A^{(2)}_T(q^2) -  \bar{A}^{(2)}_T(q^2)  \Big) ~.
\eea

Fig.~\ref{fig:AT2gAgVgP} shows $\langle A_T^{(2)} (q^2)\rangle$  for $\BDstartaunu$ in the presence of only the $g_A$ , only the $g_V $ and only the $g_P $ couplings. We make the following observations
\begin{itemize}

\item If only the $g_A$ coupling is present, the magnitude of  $\langle A_T^{(2)}(q^2) \rangle$ can go up to 10\% at  low $q^2 $ and this TP vanishes at the end points.
It can have either sign at   both low and high $q^2$. Also $\langle A_T^{(2)} \rangle$ may or may not have non-SM zero crossing.

 \item If only the $g_V$ coupling is present,  the magnitude of  $\langle A_T^{(2)}(q^2) \rangle$ can reach  up to  10\% at low  $q^2$. The behavior of $\langle A_T^{(2)}(q^2) \rangle$  is similar to the one when only the $g_A$ coupling is present.

\item If only the $g_P$ coupling is present, the asymmetry prediction is similar to the other two cases.
\end{itemize}

\FIGURE[t]{
\includegraphics[width=0.4\linewidth]{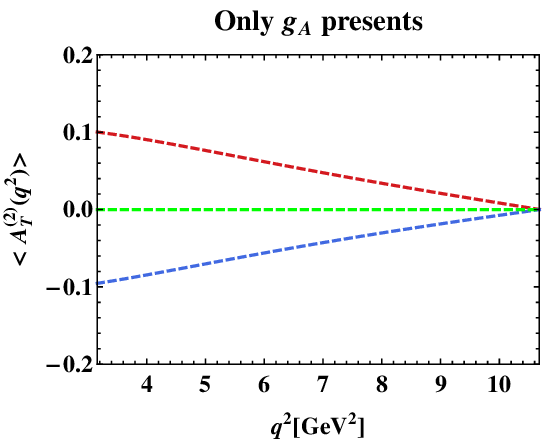}
\includegraphics[width=0.4\linewidth]{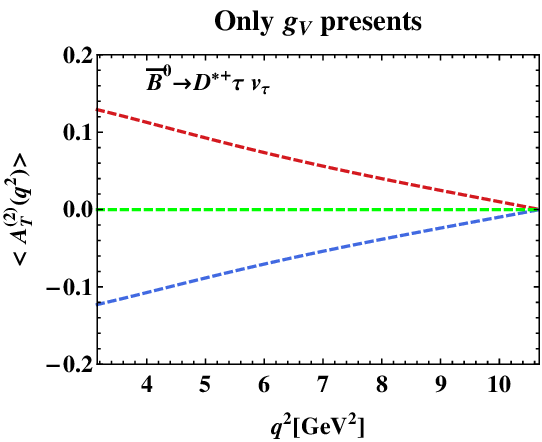}\\
\includegraphics[width=0.4\linewidth]{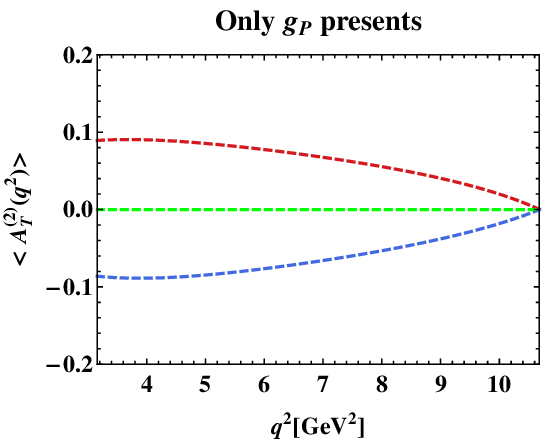}
 \caption{The figures show   $\langle A_T^{(2)}(q^2) \rangle$ for the decay $\bar{B}^0 \to D^{*+} \tau \nu_\tau$ in the scenario   where only the $g_A$ , only the $g_V$, and only the $g_P$  couplings  are present.  The green dashed line corresponds to the SM prediction. The red  and blue dashed lines correspond to $|g_A| e^{i \phi_{g_A}} = 1.32 e^{-i 0.88} $ and $|g_A| e^{i \phi_{g_A}} = 0.92 e^{i 0.27} $ respectively, in the upper-left panel,  $|g_V| e^{i \phi_{g_V}} = 1.51 e^{-i 2.11} $ and $|g_V| e^{i \phi_{g_V}} = 1.51 e^{i 2.08} $ in the upper-right panel, and  $|g_P| e^{i \phi_{g_P}} = 3.57 e^{-i 1.14} $ and $|g_P| e^{i \phi_{g_P}} = 2.86 e^{i 0.96} $  in the lower panel. The values of the couplings are chosen to show the maximum and minimum deviations from the SM expectations.   
\label{fig:AT2gAgVgP}}
}

\subsubsection{$A_T^{(3)}$}
The third TP is $A_{T}^{(3)}$, introduced above in Eq.~(\ref{eq2:Gam3}). $A_{T}^{(3)}$  and $\bar{A}_{T}^{(3)}$ are given by
\bea
 \label{eq1:AT3}
 A^{(3)}_T(q^2) &=& \frac{ V^{0T}_4}{(A_L+ A_T)},\quad \bar{A}^{(3)}_T = - \frac{ \bar{V}^{0T}_4}{(\bar{A}_L + \bar{A}_T)} \; . 
\eea
We observe that $A^{(3)}_T$ depends  on both the  new couplings $g_A$ and $g_V$ but does not depend on $g_P$. The
CP-violating triple-product asymmetry is
\bea
\label{eq2:AT3}
\langle A_{T}^{(3)} (q^2)\rangle &=& \frac{1}{2} \Big( A^{(3)}_T(q^2) + \bar{A}^{(3)}_T(q^2)  \Big) ~.
\eea 

Fig.~\ref{fig:AT3gAgV} shows $\langle A_T^{(3)} \rangle$  for $\BDstartaunu $ in the presence of only the $g_A$ and only the $g_V $ couplings. We make the following observations:
\begin{itemize}

\item If only the $g_A$ coupling is present, the magnitude of  $\langle A_T^{(3)}(q^2) \rangle$ can be  enhanced up to 4\% at  $q^2 \approx 8.0 \mathrm{GeV}^2$ and it vanishes at the end points.  $\langle A_T^{(3)} (q^2)\rangle$ can have either sign at   both low and high $q^2$. Also it may or may not have a non-SM zero crossing. 

 \item If only the $g_V$ coupling is present, the magnitude of  $\langle A_T^{(3)}(q^2) \rangle$ can be  enhanced up to 5\% at  $q^2 \approx 8.0 \mathrm{GeV}^2$.  The behavior of  $\langle A_T^{(3)} (q^2)\rangle$ is similar to the case above.
\end{itemize}

\FIGURE[t]{
\includegraphics[width=0.4\linewidth]{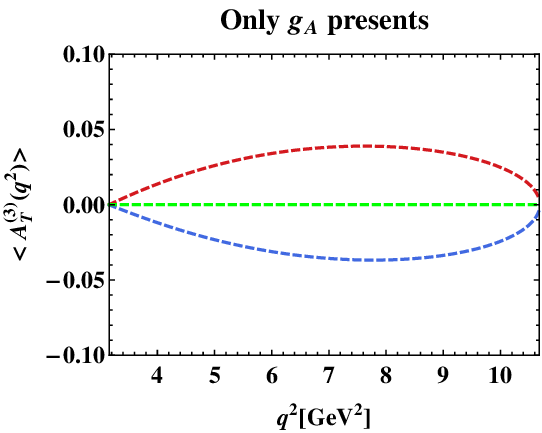}
\includegraphics[width=0.4\linewidth]{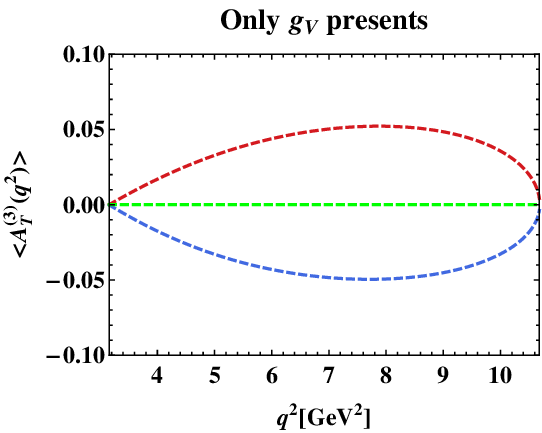}
 \caption{The left and right panels of the figure show  $\langle A_T^{(3)} (q^2)\rangle$ for the decay $\BDstartaunu$ in the scenario   where only the $g_A$  and only the $g_V$  couplings  are present. The green  dashed line corresponds to the SM prediction. The red  and blue dashed lines correspond to $|g_A| e^{i \phi_{g_A}} = 1.44 e^{-i 0.82} $ and $|g_A| e^{i \phi_{g_A}} = 1.66 e^{-i 0.74} $ respectively, in the left panel, and  $|g_V| e^{i \phi_{g_V}} = 1.75 e^{-i 2.25} $ and $|g_V| e^{i \phi_{g_V}} = 1.5 e^{i 2.1} $ in the right panel. The values of the couplings are chosen to show the maximum and minimum deviations from the SM expectations.  
\label{fig:AT3gAgV}}
}
\subsection{Correlations between $R_{D^*}$ and $q^2$-integrated TP  asymmetries}

As we discussed in the previous section, the three CP-violating TP asymmetries $
A_T^{(1,2,3)}(q^2)$ are sensitive to the new  $g_A$ and  $g_V $ couplings. It is useful to study the correlations between the $q^2$-integrated  TP asymmetries $ \langle A_T^{(1,2,3)} \rangle$ and   $R_{D^*}$ in the presence of these new couplings. Fig.~\ref{fig:qsqIntAT123} shows  the correlation between   $ \langle A_T^{(1,2,3)} \rangle$ and   $R_{D^*}$ in the presence of only the $g_A$ and only the  $g_V $ couplings. The orange(blue) color scatter plots correspond to only the $g_A(g_V)$ couplings. Here we have varied the magnitude of $g_A(g_V)$ between $(0, 2.5)$ and  its phase between $(-\pi, \pi)$. All other theoretical inputs are kept at their central values. The  vertical bands correspond to the  measured $R_{D^*}$ in Eq.(\ref{babarnew}) with $\pm 1 \sigma$ (green) and $\pm 2 \sigma$ (yellow) errors. We make the following observations for the measured $R_{D^*}$ within   $\pm 2 \sigma$  errors :
\begin{itemize}

\item If only the $g_A$ coupling is present, the magnitude of  $\langle A_T^{(1,3)} \rangle$ can   reach up to 3\% while $\langle A_T^{(2)} \rangle$ can be up to 5\%. All these asymmetries can have either sign. 

 \item If only the $g_V$ coupling is present, the magnitude of  $\langle A_T^{(1,3)} \rangle$ can  reach up to 5\% while $\langle A_T^{(2)} \rangle$ can reach up to 10\%. All these asymmetries can have either sign.
\end{itemize}

\FIGURE[t]{
\includegraphics[width=0.4\linewidth]{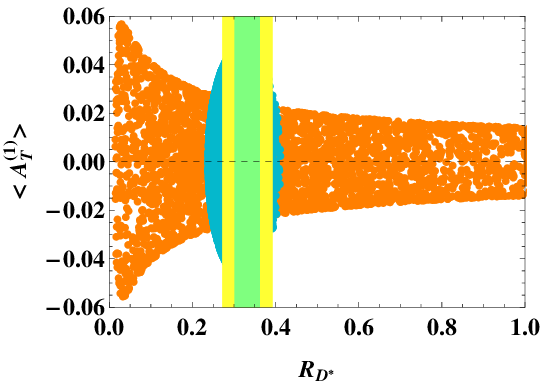}
\includegraphics[width=0.4\linewidth]{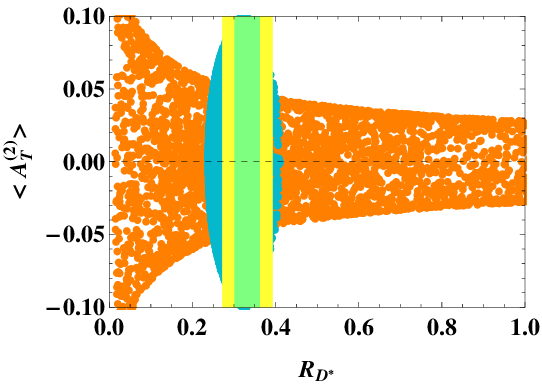}\\
\includegraphics[width=0.4\linewidth]{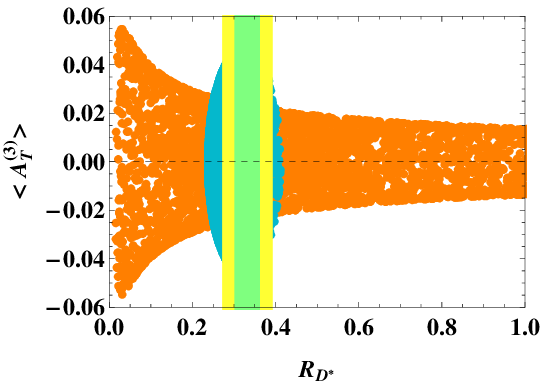}
 \caption{The figure shows the correlation between   $ \langle A_T^{(1,2,3)} \rangle$ and   $R_{D^*}$ in the presence of only the $g_A$ and only the $g_V $ couplings. See the text for  details. 
\label{fig:qsqIntAT123}}
}

The new $g_P$ coupling can only significantly affect $ \langle A_T^{(2)} \rangle$.
As shown in the pink scatter plot in  Fig.~\ref{fig:qsqIntAT2gP}, $ \langle A_T^{(2)}
\rangle$ can reach  up to 6\% for the measured $R_{D^*}$ within   $\pm 2 \sigma$  errors
in this scenario. This asymmetry can have either sign.

\FIGURE[t]{
\includegraphics[width=0.6\linewidth]{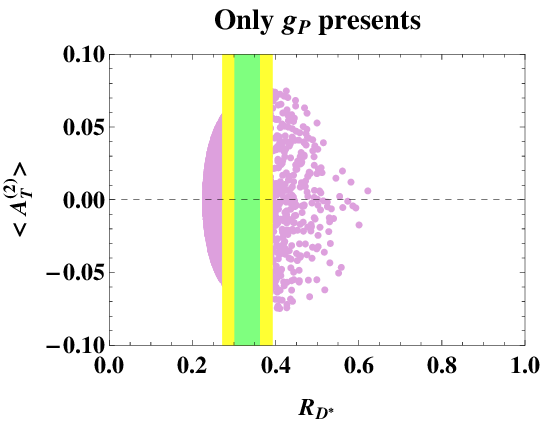}
 \caption{The figure show the correlation between   $ \langle A_T^{(1,2,3)} \rangle$ and   $R_{D^*}$ in the presence of only the $g_P$  coupling.  
\label{fig:qsqIntAT2gP}}
}

\section{Discussion and Summary}
\label{summary}
We presented 
the full three angle and $q^2$ distribution
for $\BDstarlnu$. We focused  on the decay $\BDstartaunu$, since the
 new experimental results  are not consistent with SM predictions.
We extended the work of Ref.~\cite{Datta:2012qk} by considering
additional observables from the angular distribution. Particular attention was paid to the CP violating triple product asymmetries. It was argued that in the SM these asymmetries vanish, to a very good approximation, and so  non-zero measurements of these asymmetries would be  smoking gun signals for new physics. Of the three triple product asymmetries two are sensitive to only
vector and axial vector new physics. Hence the triple product asymmetries are not only sensitive to new physics but also can probe the nature of new physics. Our results are summarized in Table~\ref{tab:summary}, for the cases where the NP has only one type of Lorentz structure: $g_A=V_R-V_L$, $g_V=V_R+V_L$, or $g_P=S_R-S_L$.

\afterpage{\clearpage}

\TABLE[!htb]{
{\footnotesize 
\begin{tabular}{p{2.6cm}|p{2.5cm}|p{2.7cm}|p{2.6cm}|p{2.6cm}}
\hline
Observable & SM & {Only new $g_A=V_R-V_L$} & {Only new $g_V=V_R+V_L$} & {Only new $g_P=S_R-S_L$} \\
\hline
 & & & & \\ 
\hfill DBR &
& $\bullet$ Significant E  
& $\bullet$  No effect  & $\bullet$ Significant E  \\
 & & & & \\ \hline
\hfill  $R_{D^*}(q^2)$ & $\bullet$ $0 \to 0.55$ \newline
(low$\to$high $q^2$)
& $\bullet$ Significant E \newline
 at high  $q^2$
& $\bullet$  No effect  & $\bullet$ Significant E \newline
 at   $q^2 \approx 7.5 \mathrm{GeV}^2$ \\ \hline
 & & & & \\
\hfill $f_L(q^2)$ & $\bullet$ $0.75 \to 0.35$ \newline
(low$\to$high $q^2$)
& $\bullet$ No effect
& $\bullet$  No effect & $\bullet$ Marginal E \\ \hline
 & & & & \\
\hfill $A_{FB}(q^2)$ & $\bullet$ ZC $\approx 5.64$ GeV$^2$ 
& $\bullet$ Significant S \newline
 at low  $q^2$ \newline $\bullet$ ZC may or may not exist
& $\bullet$ Significant S \newline
 at low  $q^2$ \newline $\bullet$ ZC may / \ may not exist  & $\bullet$ Significant E/\ S 
 at low  $q^2$ \newline $\bullet$ ZC may / \ may not exist \\
 & & & & \\ \hline
\hfill $A_C^{(1)}(q^2)$ &  $\bullet$ $0.0 \to -0.2$ \newline
(low$\to$high $q^2$)\newline $\bullet$ No ZC
& $\bullet$ No effect 
& $\bullet$ No effect  & $\bullet$ Marginal E \newline
 at   $q^2 \approx 8.0 \mathrm{GeV}^2$ \\
 & & & & \\ \hline
\hfill $A_C^{(2)}(q^2)$ & $\bullet$ $0.3 \to 0.0$ \newline
(low$\to$high $q^2$)\newline $\bullet$ No ZC
& $\bullet$ Significant S \newline $\bullet$
 ZC may / \ may not exist
& $\bullet$ Significant S \newline $\bullet$
 ZC may / \ may not exist & $\bullet$ Significant S \newline $\bullet$
 ZC may / \ may not exist \\ \hline
\hfill $A_C^{(3)}(q^2)$ & $\bullet$ $0.0 \to 0.15$ \newline
(low$\to$high $q^2$)\newline $\bullet$ No ZC
& $\bullet$ No effect
& $\bullet$ No effect & $\bullet$ Marginal S \\ \hline
\hfill $A_T^{(1)}(q^2)$ & 
& $\bullet$ Significant E/\ S 
 at $q^2 \approx 8.0 \mathrm{GeV}^2$ \newline $\bullet$
 ZC may / \ may not exist
&  $\bullet$ Significant E/\ S 
 at $q^2 \approx 8.0 \mathrm{GeV}^2$ \newline $\bullet$
 ZC may / \ may not exist & $\bullet$ No effect \\ \hline
\hfill $A_T^{(2)}(q^2)$ & 
& $\bullet$ Significant E/\ S 
 at low $q^2 $ \newline $\bullet$
 ZC may / \ may not exist
& $\bullet$ Significant E/\ S 
 at low $q^2 $ \newline $\bullet$
 ZC may / \ may not exist & $\bullet$ Significant E/\ S 
 at low $q^2 $ \newline $\bullet$
 ZC may / \ may not exist \\ \hline
\hfill $A_T^{(3)}(q^2)$ & 
& $\bullet$ Significant E/\ S 
 at $q^2 \approx 8.0 \mathrm{GeV}^2$ \newline $\bullet$
 ZC may / \ may not exist
& $\bullet$ Significant E/\ S 
 at $q^2 \approx 8.0 \mathrm{GeV}^2$ \newline $\bullet$
 ZC may / \ may not exist & $\bullet$ No effect \\
\hline
\end{tabular}
}
\caption{The effect of NP couplings on observables.
E: enhancement,
S: suppression,
ZC: zero crossing.
\label{tab:summary}}
}

\section*{Acknowledgements} This work was supported in part by the National Science
Foundation under Grant No.\ NSF PHY-1068052. AD thanks A. Soffer, M. Roney and Manuel Franco Sevilla for useful discussions.

\appendix
\section{\boldmath Details of the $\bar{B}\rightarrow D^{*} \tau^- \bar{\nu}_\tau$ angular analysis} 
\label{app-bkstarmumu}

\subsection{kinematics}

In the $B$ rest frame, the co-ordinates are chosen  such that  the $D^{*}$ meson is moving along the positive z-axis, whereas the virtual gauge boson is moving along the negative z-axis.  The  four-momenta of the B and $D^{*}$ mesons, and  the virtual gauge boson   are
\bea
\label{eqAppK2:MomB}
p_B &=& (m_B,0,0,0)\,,~~p_{D^{*}} = (E_{D^{*}},0,0,|p_{D^{*}}|) \,,~~q = (q_0,0,0, -|p_{D^{*}}|)\,,
\eea
where $E_{D^{*}} = (m^2_B + m^2_{D^{*}} -q^2)/ 2 m_B$, $|p_{D^{*}}| = \lambda^{1/2}(m^2_B,m^2_{D^{*}},q^2)/2 m_B$, and  $q_0 = (m^2_B - m^2_{D^*} + q^2)/2 m_B$.  Further, one chooses the polarization vector of the $D^*$ meson as
\bea
\label{eqAppK3:hlvecDstB}
\epsilon(0) &=& \frac{1}{m_{D^*}}(|p_{D^*}|,0,0, E_{D^*})\,,\quad
\epsilon(\pm)= \mp \frac{1}{\sqrt{2}}(0, 1, \pm i ,0)\,.
\eea

In this frame, we choose  the polarization vector of the virtual gauge boson $\bar{\epsilon}$, which can be, longitudinal $(m = 0)$, transverse $(m = \pm)$, or timelike (m = t):
\bea
\label{eqAppK1:hlvecB}
\bar{\epsilon}(0) &=& \frac{1}{\sqrt{q^2}}(|p_{D^{(*)}}|,0,0,-q_0)\,,\quad 
\bar{\epsilon}(\pm) =  \frac{1}{\sqrt{2}}(0,\pm 1, -i ,0)\,,\nl
\bar{\epsilon}(t) &=& \frac{q^\mu}{\sqrt{q^2}} =  \frac{1}{\sqrt{q^2}}(q_0,0,0,-|p_{D^{*}}|)\,,
\eea

The leptonic tensor  is evaluated in the $q^2$ rest frame. In this frame, we choose  the  transverse components of the helicity basis $\bar{\epsilon}$  to remain the same and other two components are taken as
\bea
\label{eqAppK4:hlvecLep}
\bar{\epsilon}(0) &=& (0,0,0,-1)\,,~
\bar{\epsilon}(t)=(1,0,0,0)\,.
\eea

Let $\theta_l$ be the angle between the  three-momenta  of  $D^{*}$ meson and  the charged lepton in the $q^2$ rest frame, and $\chi$ be the opening angle between the two decay planes. We define the momenta of the lepton and anti-neutrino  pairs as
\bea
\label{eqAppK5:momLep}
p^\mu_l &=& (E_l, pl \sin{\theta_l} \cos{\chi}, pl \sin{\theta_l} \sin{\chi}, -pl \cos{\theta_l})\,,\nl
p^\mu_\nu &=& (p_l, -pl \sin{\theta_l} \cos{\chi},-pl \sin{\theta_l} \sin{\chi}, pl \cos{\theta_l})\,,
\eea
where the lepton energy $E_l = (q^2 + m^2_l)/2 \sqrt{q^2}$ and the magnitude of its three-momenta  is $p_l =  (q^2 - m^2_l)/2 \sqrt{q^2}$.

\subsection{Form Factors}
The relevant form factors for the $B \to D^*$ matrix elements of the vector $V_\mu = \bar{c}\gamma^{\mu}b$ and  axial-vector  $A_\mu = \bar{c}\gamma^{\mu} \gamma_5 b$ currents are defined as \cite{Beneke:2000wa}
\bea
\label{eqApp3:MEVFF}
  \bra{ D^*(p_{D^*},\epsilon^*)}V_\mu \ket{\bar{B}(p_B)}  &=&
 \frac{2 i V(q^2)}{m_B + m_{D^*}}\varepsilon_{\mu \nu \rho \sigma} \epsilon^{*\nu}  p^{\rho}_{D^*} p^{\sigma}_B \,,\nl 
 \bra{ D^*(p_{D^*},\epsilon^*)}A_\mu \ket{\bar{B}(p_B)}  &=&  2 m_{D^*} A_0 (q^2)\frac{\epsilon^*.q}{q^2} q_\mu + (m_B + m_{D^*}) A_1(q^2) \Big[\epsilon^*_{\mu}-\frac{\epsilon^*.q}{q^2} q_\mu \Big]\nl && \hspace*{-4.5cm}-A_2(q^2) \frac{\epsilon^*.q}{(m_B + m_{D^*})} \Big[(p_B +p_{D^*})_\mu -\frac{m^2_B-m^2_{D^*}}{q^2}q_\mu \Big]\,.
\eea

In addition,  from  Eq.~(\ref{eqApp3:MEVFF}) one can show that  the $B \to D^*$ matrix element for the scalar current vanishes and for the pseudoscalar current reduces to

\bea
\label{eqApp3:MEVFF2}
\bra{ D^*(p_{D^*},\epsilon^*)}\bar{c}\gamma_5 b\ket{\bar{B}(p_B)}  &=& -\frac{ 2 m_{D^*} A_0 (q^2)}{m_b(\mu) + m_c(\mu)}\epsilon^*.q\,.
\eea


\section{ Form factors in the Heavy Quark Effective Theory}
\label{FF}
In the heavy quark limit for the b, c quarks $(m_{b,c} \gg \Lambda_{QCD})$, both charm
and the bottom quark in the hadronic current have to be replaced by static quarks  $h_{v^\prime,c}$ and $h_{v,b}$, where $v^\mu_B = p_B/m_B$ and $v^{\prime \mu}_{D^{*}} = p_{D^{*}}/m_{D^{*}} $ are the four-velocities of the B and $D^*$ mesons, respectively. The $ b \to c $ transition can be studied in the heavy quark effective theory (HQET). 
In this effective theory, the matrix elements of the vector and axial vector currents, $V_\mu$ 
and $A_\mu$ , between bottom and charm mesons \cite{Falk:1992wt} are defined as
\bea
\label{eqApp5:MEVFFHL}
  \langle D(v') |\,V_\mu\,| B(v) \rangle &=&
    \sqrt{m_B m_D}\,\Big[ h_+(w)\,(v+v')_\mu  +
    h_-(w)\,(v-v')_\mu \Big] \,, \nonumber\\
\langle D^*(v',\epsilon') |\,V_\mu\,| B(v) \rangle &=&
    i \sqrt{m_B m_{D^*}}\,\,h_V(w)\,
    \epsilon_{\mu\nu\alpha\beta}
    \,\epsilon'^{*\nu}\,v'^\alpha\,v^\beta \,, \nonumber\\
   && \nonumber\\
   \langle D^*(v',\epsilon') |\,A_\mu\,| B(v) \rangle &=&
    \sqrt{m_B m_{D^*}}\,\Big[ h_{A_1}(w)\,(w+1)\,
    \epsilon_\mu'^*  - h_{A_2}(w)\,\epsilon'^*\!\!\cdot\! v\,v_\mu
   \nonumber\\
   && - h_{A_3}(w)\,\epsilon'^*\!\!\cdot\! v\,v'_\mu \Big] \,,
\eea
where the  kinematical variable $w =v_B.v_{D^{*}} = (m^2_B + m^2_{D^{*}}-q^2)/2 m_B m_{D^{*}}$. 

The form factors $h_{A_i}(w)$ are related to the form factors in Eq.~(\ref{eqApp3:MEVFF})  \cite{ Fajfer:2012vx, Caprini:1997mu,Dungel:2010uk} in the following way,
\bea
\label{eqApp6:MEVFFHL2}
A_1(q^2) &= & R_{D^*} \frac{w+1}{2}h_{A_1}(w)\,,\quad
A_0(q^2) = \frac{R_0(w)}{R_{D^*}}h_{A_1}(w)\,,\nl
A_2(q^2) &= & \frac{R_2(w)}{R_{D^*}}h_{A_1}(w)\,,\quad
V(q^2)= \frac{R_1(w)}{R_{D^*}}h_{A_1}(w)\,,
\eea
where $R_{D^*} = 2 \sqrt{m_B m_D^*}/(m_B + m_D^*)$. The $w$ dependence of the form factors can be found in \cite{Fajfer:2012vx, Caprini:1997mu} and the summary of the results are
\bea
\label{eqApp7:MEVFFHL3}
h_{A_1}(w) &=& h_{A_1}(1)\Big[1-8 \rho^2 z + (53 \rho^2-15)z^2  -(231 \rho^2 -91) z^3\Big]\,,\nl
R_1(w)  &=& R_1(1) - 0.12 (w-1) + 0.05 (w-1)^2 \,,\nl
R_2(w)  &=& R_2(1) + 0.11 (w-1) - 0.06(w-1)^2 \,,\nl
R_0(w)  &=& R_0(1) - 0.11 (w-1) + 0.01(w-1)^2 \,,
\eea
where $z =( \sqrt{w+1}-\sqrt{2})/( \sqrt{w+1}+\sqrt{2})$. The numerical values of the free parameters $\rho^2$, $h_{A_1}(1)$, $R_1(1)$ and $R_2(1) $ are taken from \cite{Dungel:2010uk},
\bea
\label{eqApp8:MEVFFHLNum}
h_{A_1}(1) |V_{cb}| &=& (34.6\pm  0.2 \pm 1.0) \times 10^{-3}\,,\nl
\rho^2 &=& 1.214 \pm 0.034 \pm 0.009\,,\nl
R_1(1) &=& 1.401\pm 0.034 \pm 0.018\,,\nl
R_2(1) &=& 0.864 \pm 0.024 \pm 0.008 \,,
\eea
and $R_0(1) = 1.14$ is taken from  \cite{Fajfer:2012vx}. In the numerical analysis, we allow  $10 \%$ uncertainties in the $R_0(1)$ value to account for higher-order corrections.

In the HQET, the transversity amplitudes of Eq.~(\ref{tran_amp}) become
\begin{align}
\label{eq:BDstAmpHL}
{\cal{A}}_0  =  &\frac{ m_B  (1 - r_*) (w + 1) \sqrt{r_*}}{ \sqrt{ (1 + r^2_* - 2 r_* w)}}h_{A_1}(w) \Big[1 + \frac{(w-1)(1-R_2(w))}{(1 - r_*)}\Big](1-g_A)\,,\nl
{\cal{A}}_\| = & m_B \sqrt{2r_*}(w + 1)h_{A_1}(w)(1 - g_A)\,,\nl
{\cal{A}}_\perp = & - m_B   \sqrt{2r_* (w^2 - 1)} h_{A_1}(w)R_1(w) (1 + g_V)\,,\nl
{\cal{A}}_{tP} = & m_B  (1 + r_*)  \sqrt{\frac{r_* (w^2 - 1)}{( 
  1 + r^2_* - 2 r_* w)}}h_{A_1}(w) R_0(w)  \Big[(1 - g_A) - \frac{m_B^2 (1 + r^2_* - 2 r_* w)}{m_l (m_b(\mu) + m_c(\mu)) } g_P \Big]\,,
\end{align}
where $r_* = m_{D^*}/m_B$, and 

\bea
\label{eq14:DDRBstAtP}
{\cal{A}}_{tP} &=& \Big({\cal{A}}_t + \frac{\sqrt{q^2}}{m_\tau} {\cal{A}}_P \Big)\,.
\eea

\subsection{Angular coefficients}

The expressions for the twelve angular coefficients $V^{\lambda}_i$ in the
$B\rightarrow D^{*}  (\rightarrow D \pi)l^- \bar{\nu}_l$ angular distribution are summarized  according to
the $D^*$ helicity combinations $\lambda_1 \lambda_2$.

The longitudinal $V^0$'s ($\lambda_1 \lambda_2 = 0 0$) are given by
\bea
\label{eq:V0}
V^0_1 &=&  2 \Big(1 +  \frac{m^2_l}{q^2} \Big) | A_0 |^2 +  \frac{4 m^2_l}{q^2} | 
  A_{tP}|^2~,\nl
V^0_2 &=& -2 \Big(1 -  \frac{m^2_l}{q^2} \Big) | A_0 |^2~,\nl
V^0_3 &=&  -8  \frac{m^2_l}{q^2} {\rm Re}[A_{tP}  A^*_0]  ~.
\eea

The transverse $V^T$'s ($\lambda_1 \lambda_2 =++,--,+-,-+$) are given by
 \bea
 \label{eq:VT}
V^T_1 &=&  \frac{1}{2} \Big(3 +  \frac{m^2_l}{q^2} \Big) \Big(|A_{\|}|^2 + |A_\perp|^2 \Big)~,\nl
V^T_2 &=&  \frac{1}{2} \Big(1 -  \frac{m^2_l}{q^2} \Big)\Big(|A_{\|}|^2 + |A_\perp|^2 \Big)~,\nl
V^T_3 &=& -4  {\rm Re}[A_{\|}  A^*_\perp] ~,\nl
V^T_4 &=& - \Big(1 -  \frac{m^2_l}{q^2} \Big)  \Big(|A_{\|}|^2 - |A_\perp|^2 \Big)~,\nl
V^T_5 &=&  \Big(1 -  \frac{m^2_l}{q^2} \Big) {\rm Im}[A_{\|}  A^*_\perp] ~.
\eea

The mixed $V^{0T}$'s ($\lambda_1 \lambda_2 = 0\pm,\pm0$) are given by
\bea
\label{eq:VLT}
V^{0T}_1 &=& \sqrt{2}  \Big(1 -  \frac{m^2_l}{q^2} \Big) {\rm Re}[A_{\|}  A^*_0] ~,\nl
V^{0T}_2 &=& 2 \sqrt{2} {\rm Re}\Big[- A_{\perp}  A^*_0 + \frac{m^2_l}{q^2}  A_{\|}  A^*_{tP} \Big]~,\nl
V^{0T}_3 &=&  2 \sqrt{2} {\rm Im}\Big[- A_{\|}  A^*_0 + \frac{m^2_l}{q^2}  A_{\perp}  A^*_{tP} \Big]~,\nl
V^{0T}_4 &=& \sqrt{2}  \Big(1 -  \frac{m^2_l}{q^2} \Big) {\rm Im}[A_{\perp}  A^*_0]~.
\eea


\end{document}